\documentclass{aastex631}

\graphicspath{{./}{figures/}}

\begin{document}

\title{Detection of $\delta$ Scuti pulsators in the eclipsing binaries observed by TESS}

\correspondingauthor{Xinghao Chen; Xu Ding; Xiaobin Zhang}
\email{chenxinghao1003@163.com; dingxu@ynao.ac.cn; xzhang@bao.ac.cn}

\author[0000-0003-3112-1967]{Xinghao Chen}
\affiliation{Yunnan Observatories, Chinese Academy of Sciences, P.O. Box 110, Kunming 650216, China}
\affiliation{Key Laboratory for Structure and Evolution of Celestial Objects, Chinese Academy of Sciences, P.O. Box 110, Kunming 650216, China}

\author{Xu Ding}
\affiliation{Yunnan Observatories, Chinese Academy of Sciences, P.O. Box 110, Kunming 650216, China}
\affiliation{Key Laboratory for Structure and Evolution of Celestial Objects, Chinese Academy of Sciences, P.O. Box 110, Kunming 650216, China}
\author{Liantao Cheng}
\affiliation{School of Opto-electronic Engineering, Zaozhuang University, Zaozhuang 277160, China}
\author[0000-0002-5164-3773]{Xiaobin Zhang}
\affiliation{Key Laboratory of Optical Astronomy, National Astronomical Observatories, Chinese Academy of Sciences, Beijing, 100012, China}
\author{Yan Li}
\affiliation{Yunnan Observatories, Chinese Academy of Sciences, P.O. Box 110, Kunming 650216, China}
\affiliation{Key Laboratory for Structure and Evolution of Celestial Objects, Chinese Academy of Sciences, P.O. Box 110, Kunming 650216, China }
\affiliation{University of Chinese Academy of Sciences, Beijing 100049, China}
\affiliation{Center for Astronomical Mega-Science, Chinese Academy of Sciences, 20A Datun Road, Chaoyang District, Beijing, 100012, China}

\author{Kaifan Ji}
\affiliation{Yunnan Observatories, Chinese Academy of Sciences, P.O. Box 110, Kunming 650216, China}

\author[0000-0003-4829-6245]{Jianping Xiong}
\affiliation{Key Laboratory of Optical Astronomy, National Astronomical Observatories, Chinese Academy of Sciences, Beijing, 100012, China}
\affiliation{University of Chinese Academy of Sciences, Beijing 100049, China}

\author{Xuzhi Li}
\affiliation{Deep Space Exploration Laboratory, Department of Astronomy, University of Science and Technology of China, Hefei, 230026, China }
\affiliation{School of Astronomy and Space Sciences, University of Science and Technology of China, Hefei, 230026, China}

\author{Changqing Luo}
\affiliation{Key Laboratory of Optical Astronomy, National Astronomical Observatories, Chinese Academy of Sciences, Beijing, 100012, China}


\begin{abstract}
Based on 2-minute cadence TESS data from sectors 1-50, we report the results of the systematic extraction of $\delta$ Scuti-type pulsations in the 6431 eclipsing binaries with orbital periods shorter than 13 days. A total number of 242 pulsators were found in those systems, including 143 new discoveries. We examined their pulsation properties based on the H-R diagram and the relationships between the dominant pulsation period $P_{\rm dom}$, orbital period $P_{\rm orb}$, and effective temperature $T_{\rm eff}$.  As a consequence, 216 targets are likely $\delta$ Scuti stars (123 new), 11 likely $\gamma$ Doradus-$\delta$ Scuti hybrid stars (8 new), 5 likely $\beta$ Cephei stars (4 new), 4 likely $\delta$ Scuti-$\gamma$ Doradus hybrid stars (3 new), 3 likely Maia stars (3 new), 2 likely pulsating red giants (1 new), and a new unclassified star.  As for the 6 new $\delta$ Scuti pulsators in eclipsing binaries with $P_{\rm orb}$ $<$ 0.65 days, we found that 3 of them significantly exceed the upper limits of the $P_{\rm dom}$/$P_{\rm orb}$ ratio. This may indicate that $P_{\rm dom}$ and $P_{\rm orb}$ are uncorrelated for them. Finally, we statistically analyzed the dominant pulsation periods of the 216 $\delta$ Scuti stars in eclipsing binaries. Those stars concentrate around 225 $\mu$Hz and the proportion of stars in the high-frequency region is significantly higher than that of single stars, which could be ascribed to the mass transfer process.

\end{abstract}


\section{Introduction} 
Eclipsing binaries with pulsating components are very attractive objects and also are becoming popular in recent decades in understanding stellar structure and evolution. The binarity provides us with accurate ways to measure the physical parameters of the system and the component stars, meanwhile the oscillations offer us significant insight into their interiors as well as opportunities to probe the physics behind the pulsation nature of stars. The pulsating stars in eclipsing binaries have been noted since the early 1970s (Tempesti 1971; Broglia \& Marin 1974, McInally \& Austin 1977). Thanks to the space missions CoRoT (Baglin et al. 2006), Kepler (Borucki et al. 2010), and TESS (Ricker et al. 2015), a large number of eclipsing binaries with pulsating components were detected (Liakos \& Niarchos 2017; Kahraman Ali{\c{c}}avu{\c{s}} et al. 2017, 2022; Gaulme \& Guzik 2019; Shi et al. 2022). Among them, there are various types of pulsating stars, such as $\delta$ Scuti stars, $\gamma$ Doradus stars, red giants, $\beta$ Cephei stars, and Maia stars. Due to the relatively short periods and large amplitudes (Aerts et al. 2010), the $\delta$ Scuti stars in eclipsing binaries are a very useful type of variable. The $\delta$ Scuti variables are a class of late A- and early F-type stars that mainly pulsate with low-order radial and nonradial modes in the range of 0.3 to 8 hours (Aerts et al. 2010). The $\delta$ Scuti-type stars in eclipsing binaries show similar pulsation characteristics to those of single stars, but the mass transfer between the two component stars makes them very distinct evolutionary histories.  Up to now,  there are about 350 $\delta$ Scuti pulsators identified in eclipsing binaries (Liakos \& Niarchos 2017; Kahraman Ali{\c{c}}avu{\c{s}} et al. 2017, 2022; Gaulme \& Guzik 2019; Shi et al. 2022).

During the last two decades, numerous studies focused on pulsating stars in eclipsing binaries have been carried out, especially for the $\delta$ Scuti variables. Kahraman et al. (2017) compared the $\delta$ Scuti-type pulsators in eclipsing binaries with the single stars and found that the binary components pulsate at shorter periods and lower amplitudes than those of single $\delta$ Scuti pulsators. Furthermore, relationships between the pulsation periods, amplitudes, and stellar parameters have been investigated (Soydugan et al. 2006; Liakos et al. 2012; Zhang et al. 2013; Kahraman Ali{\c{c}}avu{\c{s}} et al. 2017; Liakos \& Niarchos 2017), where the well-known correlation is between the dominant pulsation period $P_{\rm dom}$ and orbital period $P_{\rm orb}$. Zhang et al.(2013) presented a theoretical explanation for this relationship, and they derived the upper limit value of the $P_{\rm dom}$/$P_{\rm orb}$ ratio for the $\delta$ Scuti pulsators in eclipsing binaries to be 0.09$\pm$0.02. Moreover, some detailed studies of the $\delta$ Scuti variables in eclipsing binaries have made attempts to understand the interiors of the stars (Guo et al. 2017; Bowman et al. 2019; Chen et al. 2020, 2021). Guo et al. (2017)  analyzed the post-mass-transfer binary KIC 9592855 and reported that the core and envelope of the pulsating component rotate nearly uniformly with its orbital motion. By combining observed properties with detailed models, Chen et al. (2020) found the Algol system KIC 10736223 had just undergone the rapid mass-transfer stage, and Chen et al.(2021) presented that OO Dra is an Algol-type binary formed through a helium-poor mass accretion. Therefore, the eclipsing binaries with pulsation components are crucial and very promising systems that offer us the opportunity to understand the intrinsic nature of pulsations and refine the theory of stellar structure and evolution.  

Asteroseismology has been included as an important research subject in the project of the new space mission China Space Station Telescope (CSST). Thus, we search for stellar pulsators in eclipsing binaries based on the TESS data to obtain a more complete database. Since a large amount of the objects are observed only once by TESS, we focus on eclipsing binary with $P_{\rm orb}$ $<$ 13 days from the catalogs of Pr{\v{s}}a et al. (2022), Shi et al. (2022), and Ding et al. (2022), and finally obtain a total of 6431 systems from the three catalogs. In this work, we aim to decipher whether there are $\delta$ Scuti variables in the TESS eclipsing binaries or not. In Section 2, we introduce the observation data and our scheme of target selection. We discuss the pulsation properties of the candidate targets in Section 3 and summarize our main results in Section 4. 

\section{Observations and target identification}
For the 6431 eclipsing binaries, we downloaded the 2-minute cadence photometric data from sectors 1-50 from the TESS Asteroseismic Science Operations Center (Tasoc, https://tasoc.dk/search\_data/).  From the files, we adopted the BJD times and the "PDCSAP\_FLUX" flux corrected for instrumental trends through the Pre-search Data Conditioning Pipeline (Jenkins et al. 2016).  Afterward,  we removed outliers and normalized the flux with the methods described in Slawson et al. (2011). For stars located in overlapped sectors, we calculated the noises of their light curves with the method in Ding et al.(2022) and adopted the data with the minimum noise for further analyses.  

For each eclipsing binary, three steps are included to test whether there is a pulsating component in the eclipsing binary or not. The details are as follows.

Firstly,  we should eliminate the changes due to eclipses from the original data. The phase of the minimum flux corresponds to the moment of the primary minimum, and the orbital period is recalculated with the method described in Ding et al. (2022) to obtain a more precise period. We computed phases and folded the light curve as illustrated in Figure 1, and then obtained the pulsation light variations by subtracting the mean curve from the folded light curve (Chen et al. 2021). In this work, 500 bins in phase are used for all the 6431 eclipsing binaries. In Figure 1, the mean light curve is shown with the red line in the top panel and the light residuals in plots of magnitude versus phase shown in the middle panel.  

Secondly, we use the Lomb-Scargle algorithm for the light residuals and show the amplitude spectrum in the bottom panel of Figure 1. Since $\delta$ Scuti stars pulsate mainly in low-order p modes with periods in the range 0.3 to 8 hours, thus we carried out further frequency analysis in the range 0-100 day$^{-1}$. In order to remove the harmonics of the orbital frequency, we generate a series of orbital-harmonic peaks n/$P_{\rm orb}$, where n varies from 1 to the integer nearest to 100/$P_{\rm orb}$. We set their amplitudes to be a fixed value (the maximum amplitude of the amplitude spectrum) and the frequency resolution 1.5$/\Delta \rm T$(Loumos \& Deeming 1978) to be the uncertainties of those orbital-harmonic peaks. We divided the amplitude spectrum by the orbital-harmonic peaks and finally obtained a pure amplitude spectrum due to intrinsic pulsations. We show the amplitude spectrum with black lines and mark the peaks that can be eliminated by the orbital harmonics with the gray lines in the bottom panel of Figure 1.  

Finally, as suggested by Breger et al. (1993) and Baran et al. (2015), the empirical threshold of the signal-to-noise ratio is S/N = 4 and 5.4 for a frequency that is accepted to be a signal, respectively. In this work, we adopt a moderate value S/N = 5 and denoted the criteria with the dotted line in the bottom panel of Figure 1. The noise is calculated in the range of 2 day$^{-1}$ around each frequency.  

We examine the pulsation amplitude spectrum for each target using the criteria of whether peaks with S/N $>$ 5 are detected or not,  and find 242 candidate $\delta$ Scuti variables in eclipsing binaries. Therein, 143 targets are new discoveries. We list their parameters in Table 1 and present their amplitude spectra in Figure 2.

\section{Properties of the pulsating stars in eclipsing binaries}
In this section, we examine the properties of the 242 candidate $\delta$ Scuti variables in eclipsing binaries. For 229 stars, there are stellar atmospheric parameters provided in the TESS input catalog, while 189 objects were also observed by Gaia (Gaia Collaboration 2022). Figures 3 and 4 present positions of the 189 targets in the Hertzsprung-Russell diagram. In the figures, the luminosities $L$ from Gaia DR3 are adopted, while the effective temperatures $T_{\rm eff}$ from the TESS input catalog are used in Figure 3 and $T_{\rm eff}$ from Gaia DR3 are used in Figure 4. It can be seen in Figure 3 that most of the targets are located in the $\delta$ Scuti instability strip except for TIC 125754991, TIC 165310952, and TIC 326356701, while many targets are beyond the blue edge of the instability strip in Figure 4.  According to the work of Uytterhoeven et al.(2011),  the typical $T_{\rm eff}$ of $\delta$ Scuti stars ranges from 6000 K to 9000 K.  TIC 326356701 with $T_{\rm eff}$ =4674 K and $\log g$=1.5484 can be a pulsating red giant.  For TIC 125754991 and TIC 165310952, we find them hotter than $\delta$ Scuti stars but cooler than $\beta$ Cephei stars and their oscillations in the range of $\delta$ Scuti stars, thus we suggest them to be two Maia variables.  

Figure 5 depicts the $P_{\rm dom}$-$P_{\rm orb}$ diagram for the 242 targets. In the figure, most of the targets manifest a consistent distribution trend with the theoretical relation derived by Zhang et al.(2013). And we also notice that 9 objects significantly exceed the upper limit of the $P_{\rm dom}$/$P_{\rm orb}$ ratio given by Zhang et al.(2013), including TIC 101531285, TIC 220511367, TIC 440459327, TIC 19193162, TIC 141622065, TIC 262958558, TIC 304030311, TIC 431555426, and TIC 449486448. In this work, we discover six new pulsating stars in eclipsing binaries with $P_{\rm orb}$ $<$ 0.65 days and display their amplitude spectra in Figure 6. As shown in Figure 6, all the six targets exhibit $\delta$ Scuti-type pulsations, which may indicate that $P_{\rm dom}$ and $P_{\rm orb}$ are uncorrelated for eclipsing binaries with $P_{\rm orb}$ $<$ 0.65 days. For TIC 19193162, TIC 141622065, TIC 262958558, TIC 304030311, TIC 431555426, and TIC 449486448, their amplitude spectra in Figure 7 suggest them to be $\gamma$ Doradus-$\delta$ Scuti hybrid stars.  

Maia stars and $\beta$ Cephei stars are two groups of pulsating stars whose frequencies resemble $\delta$ Scuti variables (Aerts et al. 2010; Balona \& Ozuyar 2020). Combining parameters in Table 1 and their amplitude spectra, TIC 315777432 may be another Maia star. Balona \& Ozuyar (2020) reported that $T_{\rm eff}$ of most $\beta$ Cephei stars are higher than 18000 K and their dominant frequencies exceed 2.5 d$^{-1}$. TIC 30562668, TIC 50897998, TIC 60433558, TIC 91111448, and TIC 335265326 are five likely $\beta$ Cephei variables. Therein, TIC 30562668, TIC 60433558, TIC 91111448, and TIC 335265326 are newly discovered stars. Besides, we suggest TIC 148560397, TIC 229742425, TIC 266006310, and TIC 339570153 to be four $\delta$ Scuti-$\gamma$ Doradus hybrid stars, and TIC 316633520, TIC 330658205, TIC 337886863, TIC 396134795, and TIC 396201681 to be five $\gamma$ Doradus-$\delta$ Scuti hybrid stars. TIC 138893145 and TIC 229771234 are two cool stars of unknown type. Gaulme \& Guzik (2019) classified TIC 138893145 to be a pulsating red giant or a triple system. TIC 229771234 is an unclassified star. Its $\log g$ from the TESS input catalog is 4.4 dex and its frequencies seem to exhibit uniformly-spaced behaviors.  

Finally, 216 targets are identified as pure $\delta$ Scuti stars. Barcel{\'o} Forteza et al. (2018) and (2020) presented a relation between the dominant frequency and the effective temperature. The distribution of the eclipsing binaries with $\delta$ Scuti variables within the $T_{\rm eff}$-$\nu_{\rm dom}$ diagram are depicted in Figure 8. In the figure, we adopt $T_{\rm eff}$ from the TESS input catalog and find that almost all targets are consistent with the relationship except for TIC 280723146. The effective temperature $T_{\rm eff}$ of TIC 280723146 from the TESS input catalog is 5711 K and 7290 K from Gaia DR3, thus we still classify it as a $\delta$ Scuti star. Figure 9 plots the histogram of the dominant pulsation frequencies of the $\delta$ Scuti stars in eclipsing binaries.  In the figure, there is a maximum value of the population of stars at 225 $\mu$Hz, which is similar to the result of $\delta$ Scuti single stars in Barcel{\'o} Forteza et al.(2020). Moreover, we found that the proportion of stars in the total number at the high-frequency region is significantly higher than that of single stars. Our result could be ascribed to the mass transfer process between the two components. Due to the mass transfer, the mass of the pulsating component increases at decreasing periods.

\section{conclusions}
In this work, we present the results of our search for pulsating components in 6431 eclipsing binaries observed by TESS in the 2-minute cadence photometric data from sectors 1-50. After eliminating the changes due to eclipses and the orbital-harmonic peaks, we found 242 binaries showing pulsation signals, of which 143 objects are new discoveries. 

Then we discuss the pulsation properties of our candidate targets based on the H-R diagram and the well-known relations between the pulsation period $\nu_{\rm dom}$ and other parameters, such as the $P_{\rm dom}$-$P_{\rm orb}$ relation and the $\nu_{\rm dom}$-$T_{\rm eff}$ relation. Among the 242 targets, 216 stars are recognized to be $\delta$ Scuti stars (123 new), 11 $\gamma$ Doradus-$\delta$ Scuti hybrid stars (8 new), 5 $\beta$ Cephei stars (4 new), 4 $\delta$ Scuti-$\gamma$ Doradus hybrid stars (3 new), 3 Maia stars (3 new), 2 pulsating red giants (1 new), and a new unclassified star. As for the six systems with $P_{\rm orb}$ $<$ 0.65 days, 3 of which significantly exceed the upper limits of the $P_{\rm dom}$/$P_{\rm orb}$ ratio. However, all the 6 targets show normal $\delta$ Scuti-type pulsations. This feature may suggest that $P_{\rm dom}$ and $P_{\rm orb}$ are uncorrelated for eclipsing binaries $P_{\rm orb}$ $<$ 0.65 days. 

We statistically analyze the dominant pulsation modes of the 216 $\delta$ Scuti stars in eclipsing binaries.  Compared with the distribution of dominant pulsation frequencies for single $\delta$ Scuti stars, there is a similar maximum value of the population of stars around 225 $\mu$Hz. Besides, we find the proportion of stars in eclipsing binaries at the high-frequency region significantly higher than that of single stars. This result may indicate that those binaries form in different mechanisms. The mass transfer makes the mean density of the mass accretors increase and its pulsation period decrease.  Such a subject will be explored in future work. 

Finally, the paper provides a vast sample of eclipsing binaries with pulsating components, which will allow for a significant improvement of the knowledge on the stellar structure and evolution and will help to understand the characteristics of pulsations in eclipsing binaries. More spectroscopic and photometric observations will be required in the future to determine their fundamental stellar parameters and constrain evolutionary models.     

\begin{acknowledgments}
We are sincerely grateful to the anonymous referee for instructive advice and productive suggestions. This work is supported by the National Key R\&D Program of China (Grant No.2021YFA1600400/2021YFA1600402) and by the B-type Strategic Priority Program No. XDB41000000 funded by the Chinese Academy of Sciences. The authors also appreciate supports of the National Natural Science Foundation of China (grant NO. 12173080 and 12288102 to X-H.C., 12103088 to X.D., 11833002 and 11973053 to X-B.Z., 12133011 and 12288102 to Y.L.). X-H.C. also sincerely acknowledges the supports of the Yunnan Revitalization Talent Support Program Young Talent Project, the Yunnan Fundamental Research Projects (202101AT070006), the science research grants from the China Manned Space Project with NO. CMS-CSST-2021-B07, and the West Light Foundation of the Chinese Academy of Sciences. L-T.C. acknowledges the support of the Natural Science Foundation of Shandong Province (ZR2021QA105). X-B.Z. also acknowledges the support of Sichuan Science and Technology Program (2020YFSY0034). X-Z.L. acknowledges the project funded by China Postdoctoral Science Foundation (2021M703099). This paper includes data collected by the TESS mission and obtained from the TESS Asteroseismic Science Operations Center (Tasoc). This paper also uses data from the European Space Agency (ESA) space mission Gaia (https://www.cosmos.esa.int/gaia), processed by the Gaia Data Processing and Analysis Consortium (DPAC, https://www.cosmos.esa.int/web/gaia/dpac/consortium). Funding for the DPAC is provided by national institutions, in particular the institutions participating in the Gaia MultiLateral Agreement. The authors sincerely acknowledge them for providing such excellent data.
\end{acknowledgments}


\begin{longrotatetable}
\begin{deluxetable*}{ccccccccccccccc}
\tablecaption{Parameters of the systems that exhibit pulsations.\label{chartable}}
\tabletypesize{\tiny}
\tablehead{
\colhead{TIC} & \colhead{R.A.} & 
\colhead{Decl.} & \colhead{$T_{\rm mag}$} & 
\colhead{$P_{\rm orb}$} & \colhead{$P_{\rm dom}$} & 
\colhead{$T_{\rm eff}^{\rm TESS}$} & \colhead{$\log g^{\rm TESS}$} & 
\colhead{$T_{\rm eff}^{\rm Gaia}$} & \colhead{$\log g^{\rm Gaia}$} & 
\colhead{$L^{Gaia}$} &\colhead{Etype} &\colhead{Ptype} &\colhead{New PB}  &\colhead{Literature} \\ 
\colhead{} & \colhead{(degree)} & \colhead{(degree)} & \colhead{(mag)} & 
\colhead{(day)} & \colhead{(day)} & \colhead{(K)} &
\colhead{(dex)} & \colhead{(K)} & \colhead{(dex)} & \colhead{($L_{\odot}$)} &\colhead{} &\colhead{} &\colhead{} &\colhead{}
} 
\startdata
              737546&       74.5150265728&      -29.9330102960& 10.5401&   3.06745&  0.058525&    7068&  3.4795&    7269&  3.6592& 27.83&EB &$\delta$ Sct &       &SHI22\\	
             3965274&       12.7806766130&      -11.1421801122& 11.5730&   0.97593&  0.017224&    8079&  4.2141&    7992&  3.8997& 11.19&EW &$\delta$ Sct &       &HOL14\\	
             7854182&      275.5486470640&       47.5688780672&  9.0619&   0.86605&  0.019993&    7424&  4.1549&    7665&  4.0699& 10.59&EB &$\delta$ Sct &       &KAH17,LIA17\\	
             8669966&      248.3713500555&       30.4990669191&  6.6493&   3.39333&  0.073412&    7651&  3.5436&        &        &      &EW &$\delta$ Sct &       &KAH22	\\
             9473243&      258.3973578307&       30.7100119195& 10.5583&   2.26697&  0.056450&    7093&  4.0218&    7553&  4.0217& 12.51&EA &$\delta$ Sct &       &KAH17,LIA17\\	
            15613315&       16.3017024552&       27.7675410431&  9.6207&   0.71665&  0.061124&    7288&  4.0251&    7251&  3.9698& 11.71&EB &$\delta$ Sct &     Y &\\	
            17336666&       65.9143623230&       21.3388350724& 10.9929&   2.95560&  0.085187&    7162&        &        &        &      &EA &$\delta$ Sct &       &SHI22\\	
            19193162&      123.7982557439&       -6.3150231583&  8.8788&   5.69178&  0.586438&    7213&  3.5239&    7286&  3.8491& 36.46&EA &$\gamma$ Dor-$\delta$ Sct &     Y &\\	
            25195864&       63.7535635637&      -69.5366740145& 10.4614&   0.74310&  0.037396&    7523&  3.7231&    8480&  3.9305& 26.06&EW &$\delta$ Sct &       &LIA17,HAS21,BAR20\\	
            27844498&      296.4906728723&       50.9058605667& 11.1924&   4.42263&  0.042329&    7722&  3.5715&    8233&  3.6902& 54.91&EB &$\delta$ Sct &       &GAU19,CUI20\\	
            30562668&      134.2814668951&      -43.2561965581&  5.9917&   3.85081&  0.114486&        &        &   18535&  3.3004&      &EB &$\beta$ Cephei &     Y &\\	
            31653503&       46.3510299657&      -66.6842455537& 10.9399&   6.68132&  0.044933&    6328&  3.4838&        &        &      &EB &$\delta$ Sct &     Y &\\	
            33834253&       63.7226832666&      -73.3604122895& 11.4218&   3.23223&  0.110378&    7140&  3.6121&    7316&  3.6851& 27.42&EA &$\delta$ Sct &     Y &\\	
            35743561&       39.8967084115&       -0.2622725567& 11.4065&   2.72087&  0.048913&    7312&  3.7668&        &        &      &EA &$\delta$ Sct &     Y &\\	
            37601240&       93.1552150914&      -25.2687352643&  8.9637&   0.73136&  0.055867&    7588&  4.2383&    7498&  4.0530&  8.28&EB &$\delta$ Sct &     Y &\\	
            37817410&       65.4180442859&       -6.0192226525&  9.2920&   2.60558&  0.053428&    7208&  3.9892&    7513&  3.8938& 13.86&EA &$\delta$ Sct &       &KAH17,LIA17,MKR18\\	
            39011225&      118.1668468136&       39.5383071699&  7.1290&   2.62881&  0.184302&    7299&  3.6066&    7263&  3.5214& 42.36&EA &$\delta$ Sct &     Y &\\	
            45294109&       92.3326604364&       18.2892200995& 13.8980&   0.36871&  0.034239&    7522&  4.2476&    7979&  4.1483& 10.96&EB &$\delta$ Sct &     Y &\\	
            45918336&      149.8182437551&      -41.2567359178& 11.2302&   2.14013&  0.059731&    8116&  4.0187&    7813&  3.9173& 15.54&EA &$\delta$ Sct &     Y &\\	
            48084398&      281.8732458071&       49.4320228548&  6.8446&   4.24361&  0.084129&    6814&  3.2927&    6862&  3.2650& 46.15&EB &$\delta$ Sct &       &SHI22,KAH22\\	
            48188257&      282.5472100339&       48.8564066458& 12.7756&   2.19167&  0.015520&    8455&  4.3322&    9826&  4.1969& 32.28&EB &$\delta$ Sct &       &KAH17,LIA17\\	
            48335665&      148.4483941173&      -34.5755893123& 10.8358&   4.62865&  0.052979&    7098&  3.5137&    7526&  3.7119& 31.50&EB &$\delta$ Sct &     Y &\\	
            49418981&      173.5889674270&       -5.5270999772&  6.4563&   2.56768&  0.081647&    7065&  3.8700&    7058&  3.8861& 12.47&EB &$\delta$ Sct &     Y &\\	
            49677785&      351.4492469005&      -11.6098951832&  9.3994&   1.59350&  0.038212&    7844&  3.8395&    7793&  3.7813& 24.94&EB &$\delta$ Sct &       &KAH17,LIA17\\	
            50897998&       83.3810301673&       -1.1560720853&  5.6058&   1.48544&  0.108923&        &        &   18266&  3.6312&      &EB &$\beta$ Cephei  &       &SOU21\\	
            54001270&      331.5410008798&      -29.5484314480& 10.6142&   2.34922&  0.041326&    7816&  3.7103&    9072&  3.9783& 54.80&EA &$\delta$ Sct &       &SHI22\\	
            54018297&       23.8996325029&      -11.9418511853&  8.8591&   1.93980&  0.075692&    7174&  3.8369&    7278&  3.8059& 16.84&EA &$\delta$ Sct &       &KAH17,LIA17\\	
            56127811&       70.7799489497&       -7.4116931399&  9.1365&   3.72351&  0.020059&        &        &        &        &      &EA &$\delta$ Sct &       &SHI22\\	
            60433558&      273.6755785367&      -33.1408936644&  8.6390&  10.80533&  0.147444&        &        &   23757&  3.8023&      &EA &$\beta$ Cephei  &     Y &\\	
            61386324&       24.2116340924&       33.6599179837&  8.3493&   0.73161&  0.050628&    7735&  4.1911&        &        &      &EB &$\delta$ Sct &     Y &\\	
            63328020&      320.0600083621&       51.3947270718& 11.5528&   1.10577&  0.049524&    7106&  3.5961&    8045&  3.7595& 37.87&EB &$\delta$ Sct &       &RAP21\\	
            64437380&      235.6983502812&      -57.5109483965&  9.8962&   2.34694&  0.033737&    7030&  3.6749&        &        &      &EB &$\delta$ Sct &     Y &\\	
            65448527&       41.9307680380&      -25.2636867300& 12.1112&   0.66781&  0.029259&    8116&  4.4009&    8109&  4.2132&  7.18&EB &$\delta$ Sct &     Y &\\	
            68032870&      219.3472146728&       38.0782749673& 10.8881&   1.03337&  0.026823&    7566&  4.0869&    7436&  3.9088& 11.58&EB &$\delta$ Sct &     Y &\\	
            68149913&      279.7729924679&       32.2562306289& 11.4431&   0.65276&  0.025305&    7790&  4.2305&    8023&  4.0912& 11.56&EB &$\delta$ Sct &     Y &\\	
            71103363&       91.9248616399&       60.8017374112& 11.8468&   2.44863&  0.020143&    8176&  4.2214&        &        &      &EA &$\delta$ Sct &     Y &\\	
            73672504&      155.5980747959&      -39.7453798220& 10.8227&   3.00204&  0.117678&    7306&  3.3123&    7284&  3.4575& 38.80&EA &$\delta$ Sct &       &SHI22\\	
            74627376&      137.9263152802&      -46.8862277885&  8.5690&   2.61257&  0.110000&        &        &    7444&  3.3223&274.90&EB &$\delta$ Sct &       &SHI22\\	
            75593781&       85.3956201214&       25.9980520192& 11.4351&   1.53207&  0.019261&    8300&  4.3274&        &        &      &EA &$\delta$ Sct &       &KAH22\\	
            78148497&       88.6019516828&       26.3087982961& 10.8176&   2.70994&  0.080356&    7287&  3.9326&    7337&  3.8903& 13.39&EB &$\delta$ Sct &       &SHI22,KAH22\\	
            79931363&      120.8536726194&      -42.1877992595&  9.7813&   0.71572&  0.035921&    7865&  4.1273&    7189&  3.8873&  8.89&EB &$\delta$ Sct &    Y &\\	
            80259373&        7.1696308623&      -44.6245691991& 10.9776&   0.63297&  0.013047&    8543&  4.5955&    9155&  4.3856&  9.59&EB &$\delta$ Sct &     Y &\\	
            81038220&      123.4914514739&       57.2659514951& 10.2772&   2.02227&  0.027026&    7526&  4.0659&    7955&  3.9565& 13.73&EA &$\delta$ Sct &       &LIA12\\	
            83833793&      217.4650186377&      -24.7242719501& 10.5126&   2.17359&  0.051926&    7469&        &    8094&  3.8313& 15.46&EA &$\delta$ Sct &     Y &\\	
            83914837&      218.0213405202&      -27.7091388555&  7.2435&   0.91181&  0.030779&    7373&  4.1943&        &        &      &EB &$\delta$ Sct &     Y &\\	
            85600400&      259.2082390246&       38.3662920672& 11.8024&   1.47224&  0.094558&    7236&  3.8548&    7239&  3.7645& 20.31&EB &$\delta$ Sct &       &KAH22\\	
            91111448&      302.0961322613&       35.4592994638&  9.1127&   1.70663&  0.075644&        &        &   18095&  4.0732&      &EB &$\beta$ Cephei  &     Y &\\	
            91369561&      312.5682800170&      -42.2179438251& 10.9545&   3.98319&  0.073957&    7105&  3.6456&        &        &      &EB &$\delta$ Sct &       &LIA17\\	
            93059429&       52.7119951153&      -13.9240825650& 11.6423&   1.48150&  0.037265&    7923&  3.9162&    8109&  3.9120& 20.33&EB &$\delta$ Sct &     Y &\\	
            95196329&      115.7817482727&      -11.1876530355& 11.7018&   0.90044&  0.064591&    7401&  3.8576&    7482&  3.8012& 23.55&EB &$\delta$ Sct &     Y &\\	
            97467902&       23.5765287783&      -27.3631260780&  7.6318&   2.04459&  0.037118&    7518&  3.7241&        &        &      &EB &$\delta$ Sct &       &SHI22\\	
           101531285&      150.2554823558&      -34.2920015279&  8.9634&   0.30363&  0.066088&    7224&  4.0989&    7130&  3.9395&  8.62&EW &$\delta$ Sct &     Y &\\	
           116334565&       85.1182206638&       30.9742414837& 11.1125&   3.47256&  0.087486&    7546&  3.5978&        &        &      &EA &$\delta$ Sct &       &KAH22\\	
           116502572&      233.8679514412&       43.4803236981& 11.3452&   3.93223&  0.061231&    6533&  3.5881&   10814&  4.1835& 79.80&EB &$\delta$ Sct &       &KAH17,LIA17\\	
           120414806&        0.1250822145&      -39.6247622077& 10.7026&   2.55280&  0.022153&    7450&  4.0020&    7813&  3.9295& 15.49&EA &$\delta$ Sct &     Y &\\	
           121078334&       56.5986922922&      -21.9721043002&  9.3429&   0.92850&  0.056547&    7421&  4.1354&    7361&  3.9847& 10.34&EB &$\delta$ Sct &     Y &\\	
           121420805&      341.8335368235&      -47.9160719778& 11.3169&   1.71701&  0.016553&    7942&  4.2016&        &        &      &EA &$\delta$ Sct &     Y &\\	
           123042728&      279.1897869675&       43.2218153400&  5.9804&   1.31265&  0.055963&    7463&  3.4943&    7315&  3.5420& 39.59&EB &$\delta$ Sct &     Y &\\	
           123042744&      279.2049048766&       43.2433834268& 11.1032&   1.31257&  0.052912&    6562&  4.2576&    6410&  4.1271&  3.30&EB &$\delta$ Sct &     Y &\\	
           124688144&       93.7211218251&      -33.0731932350& 10.6762&   1.24618&  0.022549&    8088&  4.3038&    8387&  4.1635& 11.12&EA &$\delta$ Sct &     Y &\\	
           125498216&      105.7125753163&      -11.7655945106&  9.6543&   0.71388&  0.022741&    8088&  4.3136&    7992&  4.0686&  9.21&EB &$\delta$ Sct &    Y &\\	
           125754991&       56.8727233330&       24.2883380526&  6.7792&   2.46116&  0.045673&    9962&  4.1123&   11109&  4.1204& 64.71&EA &Maia          &    Y & \\	
           126137581&      110.2637797680&       25.6688239111&  8.5546&   1.04775&  0.028804&    7723&  4.2911&    7578&  4.1499&  7.67&EB &$\delta$ Sct &     Y &\\	
           126321739&      134.9424464858&       25.8552012732& 12.7968&   0.79195&  0.028574&    7976&  4.2233&    8317&  4.1161& 17.86&EB &$\delta$ Sct &     Y &\\	
           126602778&      314.5087578290&      -46.0667710152& 10.8078&   0.71213&  0.057899&    7454&  4.0993&    7522&  4.0053& 12.19&EB &$\delta$ Sct &     Y &\\	
           126945917&      318.7323618732&      -43.3954476169&  9.1063&   1.34107&  0.045304&    7580&  3.9855&    7466&  3.9221& 14.75&EA &$\delta$ Sct &       &SHI22\\	
           129764561&       35.4020502946&      -37.2126901064&  9.9395&   2.43471&  0.039613&    7527&  3.8513&   10216&  4.1493& 47.34&EA &$\delta$ Sct &     Y &\\	
           129861919&       55.5501519986&      -38.4968177266& 10.3862&   0.70142&  0.050858&    6634&        &        &        &      &EB &$\delta$ Sct &     Y &\\	
           138893145&      294.9363790779&       39.8524649894& 10.1526&   3.75067&  0.162226&    4580&        &        &        &      &EB &RGB or Triple &       &GAU13,  GAU19\\	
           139701741&       66.3967470206&      -18.8005583916& 11.6864&   2.62979&  0.059128&    7117&  3.8023&    7677&  3.9156& 13.24&EB &$\delta$ Sct &       &SHI22\\	
           141622065&       89.2287416960&      -73.3426393715&  9.7993&   2.92792&  0.377136&    6640&  3.4682&    6861&  3.6511& 22.56&EB &$\gamma$ Dor-$\delta$ Sct &     Y &\\	
           144462697&      336.3483090293&      -44.5942263579& 11.8315&   1.23401&  0.043803&    7337&  4.2078&    7561&  4.1148&  8.00&EA &$\delta$ Sct &    Y &\\	
           147201138&      322.9532248197&      -45.0450932242& 10.9686&   1.88051&  0.030262&    7551&  3.8885&    8114&  3.9060& 21.96&EA &$\delta$ Sct &       &BOW19\\	
           148559140&      164.2816735605&       -6.2733533494&  9.7805&   0.82563&  0.023435&    7891&  4.2133&        &        &      &EB &$\delta$ Sct &     Y &\\	
           148560397&      164.3776019993&      -11.0879926170&  8.9136&   1.35017&  0.056870&        &        &        &        &      &EA &$\delta$ Sct-$\gamma$ Dor &     Y &\\	
           149160359&       81.4550908276&      -64.9893744768&  9.9927&   1.12082&  0.036331&    7931&  3.9159&    7894&  3.9102& 20.34&EB &$\delta$ Sct &       &WK20\\	
           150443185&       97.8678805718&      -63.5353762050&  9.4343&   2.59376&  0.021750&    7421&  3.9282&    7500&  3.9310& 16.46&EA &$\delta$ Sct &     Y &\\	
           152223725&      312.2794999388&      -33.7317933625&  9.3148&   4.43644&  0.081727&    7708&  3.4433&    9390&  3.7611&113.80&EA &$\delta$ Sct &       &KAH17,LIA17\\	
           152513129&      172.1072965932&      -39.7298573935& 12.0743&   4.12675&  0.020192&    7575&  4.0418&    9842&  4.2509& 26.82&EB &$\delta$ Sct &     Y &\\	
           157273321&      198.7268625502&       59.2956675972&  7.8863&   5.51960&  0.059238&    7285&  3.5397&   10352&  3.8798&111.00&EB &$\delta$ Sct &       &KAH17,LIA17\\	
           158794976&      288.7183567756&       42.5022843106&  9.9366&   0.77267&  0.036402&    7510&  4.0154&    7462&  3.9550& 12.82&EW &$\delta$ Sct &       &SHI22\\	
           159298033&      229.4895851838&       83.8594649404&  7.3670&   1.72499&  0.055414&    7204&  4.0067&    7329&  3.8972& 11.88&EB &$\delta$ Sct &     Y &\\	
           159638876&      317.7559917720&      -40.5393026742&  8.1990&   2.39299&  0.091245&    7089&  3.5983&    6956&  3.6509& 24.36&EB &$\delta$ Sct &     Y &\\	
           159864583&      221.4063516221&      -40.0303761311&  8.5822&   3.63856&  0.162635&    6991&  3.8801&    6885&  3.8322& 11.83&EB &$\delta$ Sct &     Y &\\	
           160036449&      336.7522882506&      -37.1883684346& 10.8022&   1.89317&  0.020405&    7517&  4.0996&        &        &      &EA &$\delta$ Sct &       &SHI22\\	
           160081043&      352.0531716144&      -39.9232510262& 12.9518&   0.76866&  0.045813&    7800&  4.3552&    7936&  4.2609&  7.26&EB &$\delta$ Sct &     Y &\\	
           163008999&      242.3037019576&      -43.1986572213&  9.6266&   1.85412&  0.106766&    6850&  3.6464&        &        &      &EA &$\delta$ Sct &     Y &\\	
           164646288&      283.9097468481&       42.1310107932& 10.5013&   0.73378&  0.042476&    7448&  4.1086&    7692&  4.0132& 12.74&EB &$\delta$ Sct &       &KAH17,LIA17\\	
           165310952&      299.2429420462&       54.7995223834&  8.4462&   1.80517&  0.131684&    9103&        &   10618&  3.7124&240.90&EB &Maia &     Y &\\	
           165459595&      180.1920401708&      -40.3545012557&  9.3150&   3.33691&  0.046010&    7697&  4.0832&    7470&  3.9615& 12.25&EA &$\delta$ Sct &       &SHI22\\	
           165618747&      260.0324962828&       13.6660148220& 11.4332&   0.64823&  0.026403&    7603&  4.1085&    8308&  4.1310& 13.96&EB &$\delta$ Sct &       &KAH22\\	
           166874908&       59.6597216407&      -31.2772956509& 10.1725&   2.18916&  0.039400&    7635&        &    7802&  3.9464& 33.75&EB &$\delta$ Sct &     Y &\\	
           169168102&      127.1214720000&       -2.5171730000&  6.1070&   2.49973&  0.029610&        &        &        &        &      &EA &$\delta$ Sct &       &BAR20\\	
           169532543&       78.0438100581&      -13.1996046163&  9.7605&   0.91541&  0.031379&    7928&  4.0117&   10051&  4.2255& 37.32&EB &$\delta$ Sct &       &KAH17,LIA17\\	
           172431974&      299.7087813125&       39.3311433479& 10.5887&   2.34477&  0.187775&    7236&  3.7116&        &        &      &EA &$\delta$ Sct &       &KAH22\\	
           175405906&      349.8565525181&      -40.2906698848& 11.3969&   2.01250&  0.054876&    6920&        &    7161&  3.8331& 18.53&EA &$\delta$ Sct &     Y &\\	
           177283493&      102.4684550367&      -73.4007811676& 11.8481&   0.75281&  0.052694&    7278&  3.9772&    7753&  3.9970& 13.97&EB &$\delta$ Sct &     Y &\\	
           178286775&       60.6982435098&      -18.8247670600& 10.5348&   1.12638&  0.018187&    7557&  4.2620&    7880&  4.0876&  8.97&EB &$\delta$ Sct &     Y &\\	
           178996712&       70.0633105020&      -27.7350809463&  8.5850&   2.98205&  0.034765&    7188&  4.1017&    7282&  4.0275&  8.81&EB &$\delta$ Sct &     Y &\\	
           181043970&       12.9441610409&      -71.9980354033& 10.4156&   0.87092&  0.017011&    8684&  4.1875&    8428&  4.2379& 14.28&EA &$\delta$ Sct &     Y &\\	
           183143120&      125.0177546163&      -38.4098114462& 11.3309&   0.49390&  0.027314&    8312&  4.3537&        &        &      &EB &$\delta$ Sct &     Y &\\	
           184246521&      295.8339922256&       39.9522626606& 10.7680&   2.47025&  0.052371&    7753&  3.5174&    7594&  3.4853& 44.17&EB &$\delta$ Sct &       &GAU19,LIA20\\	
           184318481&      127.0757311805&      -38.9720322442&  8.4729&   1.55787&  0.021267&    7899&  4.3156&    8077&  4.3044&  9.02&EA &$\delta$ Sct &     Y &\\	
           188630094&      350.5857754196&      -15.9390248074& 10.4939&   0.86275&  0.030981&    7354&  3.9536&    7415&  3.9156& 12.94&EB &$\delta$ Sct &       &KAH17,LIA17\\	
           193391232&      305.7618343832&       43.2423563160& 10.4713&   3.45091&  0.031176&    7681&  3.8612&        &        &      &EA &$\delta$ Sct &       &KAH17,LIA17\\	
           193774939&      268.3030649995&       43.7731023005&  9.8556&   1.30579&  0.046307&    7358&  4.0550&    7506&  3.9778& 11.47&EA &$\delta$ Sct &       &KAH22\\	
           197757000&      337.2080240079&       53.7710780374& 10.4189&   2.20651&  0.027283&    6985&  3.9230&   10057&  4.2701& 36.06&EA &$\delta$ Sct &       &KAH22\\	
           198037741&       61.6537070111&      -56.5752786716&  8.5467&   0.78000&  0.048889&    7274&  3.8397&    7463&  3.9550& 12.93&EB &$\delta$ Sct &     Y &\\	
           199688409&      253.0514502264&       57.7254690748& 12.0490&   0.86811&  0.033204&    7738&  4.1345&    7824&  4.0069& 13.03&EB &$\delta$ Sct &     Y &\\	
           200440270&       70.8966718559&      -42.6758227821& 10.1114&   1.07885&  0.016486&    8123&  4.1742&    9793&  4.1923& 24.17&EB &$\delta$ Sct &       &BAR20\\	
           202299805&      225.1608428572&       57.7751470014& 11.5680&   1.54580&  0.018534&    7709&  4.2172&    7954&  4.0748&  9.24&EA &$\delta$ Sct &     Y &\\	
           214330573&      261.1089384280&      -52.5672613424&  9.2178&   1.91207&  0.092655&    6662&  3.5863&        &        &      &EB &$\delta$ Sct &     Y &\\	
           217279153&      260.1266540055&      -43.0762219588& 10.6201&   1.60971&  0.042121&    7216&  3.8748&    7745&  3.9741& 15.20&EA &$\delta$ Sct &       &LIA17\\	
           218996233&      124.7665718848&      -18.5659931911& 11.6094&   0.77637&  0.040478&    7551&  4.3347&    7925&  4.1870&  7.15&EA &$\delta$ Sct &     Y &\\	
           219373406&       76.2354874222&      -53.1405873748& 10.5108&   0.86189&  0.019870&    7766&  4.0242&    7933&  4.0459& 15.02&EB &$\delta$ Sct &     Y &\\	
           219707463&      136.5934902441&       68.4451458176&  6.8367&   9.01513&  0.045098&    7541&  3.9818&        &        &      &EA &$\delta$ Sct &       &SHI22\\	
           220402294&       69.5598201448&      -57.1921582692& 11.2216&   0.78521&  0.028387&    7258&  4.0560&    7279&  3.9845& 10.30&EB &$\delta$ Sct &     Y &\\	
           220511367&      347.4854084181&      -54.6222138491&  9.5287&   0.31750&  0.046722&    7566&  4.3053&    7489&  4.1765&  6.53&EW &$\delta$ Sct &     Y &\\	
           220962774&      246.5646776038&      -55.2088921324&  9.2632&   0.99449&  0.047819&    6821&  3.8944&    7318&  3.9436& 13.43&EB &$\delta$ Sct &     Y &\\	
           229742425&      279.7204436093&       70.2315768197& 10.3807&   1.26528&  0.049903&    7260&  4.1469&    7364&  4.0851&  8.44&EB &$\delta$ Sct-$\gamma$ Dor&     Y &\\	
           229771234&      282.2677475537&       68.1569510253& 10.7665&   0.82092&  0.047984&    4000&  4.4290&        &        &      &EB &unclassified       &Y &\\	
           231714759&      352.5927028842&      -58.4262810043&  9.3827&   5.46147&  0.030842&    7087&  3.6567&    7571&  3.7370& 26.12&EB &$\delta$ Sct &     Y &\\	
           231973885&       95.5053539497&      -47.8992013797& 11.0410&   2.89466&  0.044452&    7545&  3.7785&    9958&  4.1059& 63.44&EA &$\delta$ Sct &       &SHI22\\	
           232052561&       97.3117988385&      -45.8704641837& 10.0244&   3.01064&  0.155260&    7256&  4.1364&    7313&  4.1212&  8.04&EA &$\delta$ Sct &     Y &\\	
           233195058&      264.2695830000&       67.1192300000&  7.7650&   2.67930&  0.048528&        &        &        &        &      &EA &$\delta$ Sct &     Y &\\	
           233291864&      300.4182617987&       58.0971758417&  9.9607&   2.68961&  0.047421&    8062&  4.2450&    7883&  4.1491& 10.18&EA &$\delta$ Sct &     Y &\\	
           238607300&      132.7133472298&       19.3572963923& 11.8142&   1.32375&  0.076777&    7114&  3.7702&    7296&  3.7998& 20.95&EB &$\delta$ Sct &       &SHI22\\	
           239291500&      300.3325178451&       44.1441255725& 11.9143&   1.61313&  0.015519&    7665&  4.0578&    9588&  4.2580& 28.28&EB &$\delta$ Sct &       &GAU19\\	
           239984394&       94.4646257585&       32.5045973943&  7.1075&   5.45998&  0.124824&    6890&  3.3870&    6870&  3.5534& 31.95&EB &$\delta$ Sct &     Y &\\	
           240947891&       18.5719675646&       48.9904537236& 10.8465&   2.49478&  0.035267&    7716&        &    9651&  4.1432&153.50&EA &$\delta$ Sct &       & KAH17,LIA17\\	
           242149417&      210.2075677412&      -42.3483150547&  8.9878&   1.82075&  0.047060&    7156&  3.7214&    7371&  3.7415& 21.52&EB &$\delta$ Sct &     Y &\\	
           243287588&      204.0331654514&      -45.2380151786& 10.4227&   2.31007&  0.084694&    6954&  3.8396&    7038&  3.8576& 14.06&EB &$\delta$ Sct &     Y &\\	
           244208023&       73.2858034467&       -3.4979848940& 11.8979&   1.82289&  0.039235&    7647&  4.0785&    9841&  4.2391& 33.00&EA &$\delta$ Sct &       &SHI22\\	
           244250449&       74.3102719895&       -3.6012937725& 10.2935&   1.74608&  0.026163&    8199&  4.0336&    9703&  4.0288& 31.79&EA &$\delta$ Sct &     Y &\\	
           255744818&       99.2595648337&      -52.0164292068&  9.2276&   3.04779&  0.044587&    7858&  3.6892&        &        &      &EB &$\delta$ Sct &     Y &\\	
           255921197&      307.4419940901&       71.4186300528&  9.0469&   1.77827&  0.053386&    6465&  3.6836&    6629&  3.7630& 13.86&EB &$\delta$ Sct &     Y &\\	
           256640561&      327.8973718744&       71.8857451308&  8.0459&   1.72079&  0.098803&    7891&  3.4821&        &        &      &EB &$\delta$ Sct &       &KAH22\\	
           258351350&      296.5105834547&       69.9191670524&  7.9352&   0.77286&  0.019205&    8015&  4.0584&    7757&  3.8228& 16.27&EB &$\delta$ Sct &       &KAH17,LIA17\\	
           262412046&       69.2764861541&        1.6919883341& 10.5627&   2.04310&  0.057038&    7070&  3.8530&        &        &      &EA &$\delta$ Sct &       &KAH17,LIA17\\	
           262958558&       11.4612395273&      -78.8213369345&  9.6435&   3.53606&  0.783152&    7616&  3.9160&    9933&  4.0202& 41.93&EA &$\gamma$ Dor-$\delta$ Sct &     Y &\\	
           264483228&       81.3797073396&        2.8833898291& 10.0242&   1.94997&  0.019339&    7485&  4.1489&    9434&  4.3881& 15.03&EA &$\delta$ Sct &     Y &\\	
           265591866&      331.9794049517&      -59.8749106526& 10.8404&   2.28081&  0.049442&    7381&  3.8815&    7504&  3.7340& 25.45&EA &$\delta$ Sct &       &SHI22\\	
           266006310&       95.3573953535&        2.2685045863&  6.0571&   5.56393&  0.050397&    7524&        &        &        &      &EA &$\delta$ Sct-$\gamma$ Dor &       &SHI22\\	
           266032748&       14.4979153234&        9.6781442865& 11.7454&   0.81256&  0.019889&    8048&  4.2999&    7600&  3.8550&  9.65&EB &$\delta$ Sct &    Y &\\	
           266735682&       29.2078661578&      -21.1954116513& 10.3677&   1.13521&  0.022483&    7828&  4.2704&    7932&  4.1815& 12.85&EA &$\delta$ Sct &     Y &\\	
           268383780&      299.0568199609&       47.9093449410& 11.3625&   2.61837&  0.043326&    8732&  4.0829&    9719&  4.0205& 42.05&EA &$\delta$ Sct &       &KAH17,LIA17\\	
           268387175&      299.0405608283&       46.6611494610& 10.8433&   2.16392&  0.096163&    6279&        &    7033&  3.5500& 35.94&EB &$\delta$ Sct &       &GAU19\\	
           268718018&       18.2608735129&      -22.6099862092& 11.7837&   0.88539&  0.018976&    7926&  4.3542&    7996&  4.1180&  8.90&EB &$\delta$ Sct &     Y &\\	
           270616815&      293.6933299764&       45.5170399445& 11.8466&   3.04445&  0.036543&    7724&  3.8304&    9230&  3.8610& 58.79&EB &$\delta$ Sct &       &GAU19\\	
           271529113&      354.5845210063&       64.3340831187&  8.9400&   2.34009&  0.031478&    7859&  3.9279&    7683&  3.8063& 21.70&EA &$\delta$ Sct &       &KAH17,LIA17\\	
           271967202&      295.9351347347&       43.4781714609& 11.6069&   3.20715&  0.035292&    7305&  3.8199&    7579&  3.8196& 21.49&EB &$\delta$ Sct &       &GAU19\\	
           274348764&      300.5163949319&       39.7656523630& 10.1646&   0.83205&  0.038466&    8166&  3.9600&    8266&  3.9565& 23.86&EB &$\delta$ Sct &     Y &\\	
           276457924&      228.4707852865&      -37.6557765624& 10.5355&   5.09962&  0.073171&    7110&  3.8320&    6983&  3.8224& 13.15&EA &$\delta$ Sct &     Y &\\	
           277565268&      311.1240403010&       54.1019402551&  9.3321&   1.09588&  0.102578&    8057&  3.7197&    7329&  3.7646& 25.78&EB &$\delta$ Sct &     Y &\\	
           280723416&      344.4304844987&      -78.4081979271& 11.4042&   0.82504&  0.016797&    5711&        &    7290&  3.4359&      &EB &$\delta$ Sct &     Y &\\	
           287351976&      122.6813197842&      -72.5457203008& 11.6313&   1.92224&  0.026258&    7436&  3.9531&    9728&  4.2309& 35.57&EA &$\delta$ Sct &       &SHI22\\	
           289722957&      274.1805028867&       53.2464959099& 11.5104&   2.21816&  0.043891&    7070&  4.0390&    7620&  4.0253& 11.52&EA &$\delta$ Sct &     Y &\\	
           291329564&      260.9457036626&      -55.0621651642&  9.9161&   2.38018&  0.045544&    7016&  3.6934&    7566&  3.7346& 26.25&EA &$\delta$ Sct &     Y &\\	
           298541138&      261.7471759963&       24.6965572269& 11.3358&   0.74908&  0.027565&    7605&  4.1217&    7898&  4.0210& 12.25&EB &$\delta$ Sct &     Y &\\	
           298734307&      278.6095797774&       57.8018197869&  7.2099&   0.94445&  0.032181&    8014&  3.9479&    7841&  3.9090& 20.62&EB &$\delta$ Sct &       &SHI22\\	
           300012172&      106.8271061047&      -66.0631983263& 10.3883&   1.46120&  0.055820&    7495&  3.8066&    7880&  3.8483& 22.95&EB &$\delta$ Sct &     Y &\\	
           300654002&      114.4168954280&      -69.5423284166& 10.2724&   2.75804&  0.051581&    7291&  3.5938&    9764&  3.9373& 73.36&EA &$\delta$ Sct &     Y &\\	
           301063488&      298.3512269214&       20.3274543770&  7.0814&   7.32720&  0.037265&    7725&        &    7115&  3.5437& 47.47&EA &$\delta$ Sct &     Y &\\	
           301405723&       52.9760428992&       -1.6392895686& 11.1728&   0.94262&  0.023323&    8089&  4.1211&    7785&  3.8965& 10.77&EA &$\delta$ Sct &       &SHI22\\	
           301407485&       53.1046925995&       -3.3133600126&  8.1191&   2.66407&  0.016941&    8053&  4.0317&   10463&  4.1744& 45.30&EB &$\delta$ Sct &       &KAH17,LIA17\\	
           302373103&      133.7731244302&      -67.5349557181& 11.5238&   0.90277&  0.017960&    7450&  4.2053&    8152&  4.0477& 11.71&EB &$\delta$ Sct &     Y &\\	
           303733089&      168.7201608338&      -34.6349217958& 10.5561&   1.26288&  0.023506&    7549&  4.0584&    8121&  3.8949& 18.62&EB &$\delta$ Sct &     Y &\\	
           304030311&      141.2246377398&      -71.4721102773& 10.4943&   6.29391&  0.730948&    7403&  4.1696&    7448&  4.0909&  8.45&EA &$\gamma$ Dor-$\delta$ Sct &     Y &\\	
           307990280&      129.2359834347&      -67.7447505240&  9.4314&   2.31598&  0.022152&    7313&  4.0282&   10812&  4.3375& 37.67&EW &$\delta$ Sct &     Y &\\	
           309658221&       79.2537140265&      -61.5145883481&  9.8567&   7.59543&  0.100535&    6813&        &        &        &      &EA &$\delta$ Sct &       &LEE19\\	
           312621168&      210.8107898102&      -48.7917478218&  9.8539&   0.93267&  0.013224&    8520&  4.3786&   10074&  4.3725& 21.14&EB &$\delta$ Sct &     Y &\\	
           312865025&      126.4657786341&       77.2185691500&  9.4026&   2.73500&  0.074584&    7103&  3.5472&    6928&  3.6346& 25.82&EA &$\delta$ Sct &     Y &\\	
           314864553&       53.1681854402&      -75.8888137155& 11.9207&   1.05562&  0.016886&    7438&  4.0832&    8378&  4.0022& 15.66&EB &$\delta$ Sct &     Y &\\	
           315777423&      321.0968862035&       56.3615613539&  8.2842&   4.93133&  0.172826&   10082&        &   17069&  3.4563&      &EA &Maia &     Y &\\	
           316633520&      263.7627392634&       24.5630259173& 11.5439&   4.64296&  0.133189&    6752&  3.3601&    6869&  3.4340& 40.47&EA &$\gamma$ Dor-$\delta$ Sct &     Y &\\	
           318088293&      307.0009082223&      -79.9952929682& 10.8986&   0.69738&  0.037317&    7684&  4.0192&    8056&  4.0373& 15.59&EB &$\delta$ Sct &     Y &\\	
           318217844&      142.8526877358&       49.4678928998&  9.9964&   1.22227&  0.020602&    7828&  4.0446&    7490&  3.8964& 16.66&EB &$\delta$ Sct &     Y &\\	
           323292655&      130.8008244196&      -79.0701017167&  5.8258&   1.66983&  0.086018&    7444&  3.6538&    7417&  3.6871& 28.61&EA &$\delta$ Sct &       &KAH17,LIA17\\	
           326356701&      325.5969830408&      -21.7626556960&  8.3387&   1.11216&  0.019724&    4674&        &    5604&  1.5484&203.00&EB &RGB    &Y &\\	
           330658205&      136.4043732507&       12.5897309304& 10.8392&   3.60217&  0.234463&    7062&  3.9792&        &        &      &EA &$\gamma$ Dor-$\delta$ Sct &       &SHI22\\	
           332541024&      270.4920720868&       30.7879611343& 10.8162&   1.17635&  0.017887&    7221&        &    7586&  3.8263& 21.90&EB &$\delta$ Sct &     Y &\\	
           335265326&      332.1899776964&       61.0224185676&  7.2771&   3.80835&  0.197503&        &        &   20267&  3.5716&      &EB &$\beta$ Cephei  &     Y &\\	
           337094559&      335.6357207166&       63.5862767539&  9.3502&   1.78066&  0.110446&    7424&  3.7162&        &        &      &EB &$\delta$ Sct &       &KAH22\\	
           337886863&      337.3230520488&       65.1133243340&  7.5536&   5.69258&  0.342182&   11686&        &        &        &      &EA &$\gamma$ Dor-$\delta$ Sct &     Y &\\	
           338159479&      338.0648451224&       64.9778629452& 11.2918&   1.41434&  0.036928&        &        &        &        &      &EB &$\delta$ Sct &       &KAH22\\	
           339397374&      295.7759845138&      -60.7787350305& 10.3223&   1.18207&  0.032584&    7367&  3.9591&    7428&  3.8894& 16.74&EB &$\delta$ Sct &     Y &\\	
           339570153&      253.6228783296&      -41.6541530122&  7.9038&   6.34726&  0.101518&        &        &        &        &      &EA &$\delta$ Sct-$\gamma$ Dor &     Y &\\	
           347361290&       25.9662025404&       10.9502834784& 10.4749&   1.01553&  0.015090&    8082&  4.2923&    9418&  4.3478& 18.35&EW &$\delta$ Sct &     Y &\\	
           350443417&       85.3046330415&      -57.4411661124&  8.9808&   1.21673&  0.043892&    7636&  4.3227&    7406&  4.1776&  6.50&EA &$\delta$ Sct &     Y &\\	
           351473399&      308.6705424632&      -64.1744980436& 11.5763&   0.86760&  0.031680&    7745&  4.2283&    7929&  4.0870& 11.05&EB &$\delta$ Sct &     Y &\\	
           351667692&      274.9476039940&       42.1772730478& 11.3446&   2.07291&  0.017925&    7939&  4.2302&    9375&  4.2222& 24.11&EB &$\delta$ Sct &     Y &\\	
           353854078&      271.1411728552&       58.3984180327&  9.8329&   5.17034&  0.043971&    6595&  3.4075&    6833&  3.4668& 31.84&EB &$\delta$ Sct &       &KAH17,LIA17\\	
           354922610&       39.3812971343&       71.3045204584&  9.8471&   1.36696&  0.058284&    7056&  4.0692&    7478&  3.9409& 10.66&EA &$\delta$ Sct &       &LIA17,HON17,MIS22\\	
           354926863&       39.4423509141&       67.8555200171& 10.5010&   1.43602&  0.046164&    7388&  4.0716&        &        &      &EA &$\delta$ Sct &       &KAH22\\	
           356194871&      266.6269381314&       53.1994297859& 10.1756&   2.71044&  0.040396&    7782&  3.7197&    9638&  3.9723& 66.61&EB &$\delta$ Sct &       &KAH17,LIA12,LIA17\\	
           356240274&      267.7113679711&       50.0471947344& 11.7539&   2.34757&  0.041004&    7394&  3.9104&   10092&  4.2015& 45.95&EA &$\delta$ Sct &     Y &\\	
           358613523&      175.1028937630&       80.2359806993& 10.0086&   1.32833&  0.024065&    7271&  4.2780&    7383&  4.1111&  6.60&EA &$\delta$ Sct &       &KAH22\\	
           363852956&      200.6195886680&      -45.8915478481& 11.8138&   0.74065&  0.060809&    7349&  4.0527&    7395&  4.0152& 10.40&EB &$\delta$ Sct &     Y &\\	
           365956683&      276.0468304400&      -63.9486439979& 10.8967&   5.73042&  0.075607&    6455&  3.4342&    6859&  3.5272& 30.38&EB &$\delta$ Sct &       &KAH17,LIA17\\	
           372759279&      175.0060004748&       75.1559630489& 10.8419&   1.23833&  0.023884&    7653&  3.9519&    7896&  3.9695& 17.76&EA &$\delta$ Sct &  &ZXB14,KAH17,LIA17,LEE18,CXH21\\	
           373029274&       67.0687637360&        7.8441154302& 11.4338&   1.07661&  0.015459&    7624&  4.2509&        &        &      &EB &$\delta$ Sct &     Y &\\	
           376486632&      180.5928829261&      -58.2272780000& 11.1646&   2.60213&  0.071493&    7422&     &           &        &      &EA &$\delta$ Sct &     Y &\\	
           378042585&      352.0751940586&       63.3908139172&  7.9317&   3.25952&  0.143783&    6965&  3.2255&    6734&  3.3628& 48.81&EB &$\delta$ Sct &     Y &\\	
           381011161&      303.7036406499&       34.7395150992& 11.9215&   1.31250&  0.030774&    7189&  3.8429&    7714&  3.9531& 18.84&EA &$\delta$ Sct &       &KAH17,LIA17\\	
           381948745&       76.3683365792&      -56.3028684329&  9.8456&   3.19170&  0.050038&    7085&  3.5597&        &        &      &EB &$\delta$ Sct &     Y &\\	
           385916950&      111.3294120580&      -16.4590270867&  8.8819&   1.56251&  0.040473&    6907&  3.7274&    6782&  3.8360& 14.58&EA &$\delta$ Sct &     Y &\\	
           388244501&      126.3858149885&      -84.3634848912&  8.5565&   2.11893&  0.048901&    7854&  3.6355&        &        &      &EB &$\delta$ Sct &     Y &\\	
           391902612&      106.5680497401&      -76.8392707703& 11.8625&   4.55234&  0.086437&    6586&  3.5690&        &        &      &EA &$\delta$ Sct &       &SHI22\\	
           393636596&      239.5767093170&       17.2694459987& 10.6639&   0.87934&  0.045744&    7568&  4.2485&    7666&  4.2340&  8.11&EA &$\delta$ Sct &       &KAH17,LIA17\\	
           393894013&      198.3890021745&       47.7977363777&  9.1556&   0.81588&  0.035586&    7866&  4.1933&    7684&  4.0579& 10.32&EB &$\delta$ Sct &       &KAH22\\	
           394126943&        2.7039836475&      -68.5472164131&  9.6443&   1.84588&  0.018387&    7610&  4.1009&    7579&  3.7353& 13.02&EB &$\delta$ Sct &     Y &\\	
           396134795&      357.0983261865&       36.3111877125&  9.5636&   2.58883&  0.070438&    6946&  3.8002&    7101&  3.7057& 19.29&EA &$\gamma$ Dor-$\delta$ Sct &       &KAH22\\	
           396201681&      359.5249573147&       67.6031600612&  9.6904&   7.03806&  0.478076&    7767&        &    6166&  2.8834& 15.15&EA &$\gamma$ Dor-$\delta$ Sct &      &KAH22\\	
           397048159&       73.3951521526&        5.6126918125& 10.1721&   1.60553&  0.063577&    7222&  3.9114&    7442&  3.8681& 14.89&EB &$\delta$ Sct &     Y &\\	
           400105121&       97.0078813768&       77.8260806041& 11.0266&   1.44492&  0.016725&    8272&  4.2594&    9626&  4.2452& 19.70&EA &$\delta$ Sct &     Y &\\	
           401701828&      317.5806712791&      -82.4240932579&  9.7728&   1.16750&  0.061965&    7428&  3.8932&    7636&  3.8679& 18.53&EB &$\delta$ Sct &     Y &\\	
           401869914&       51.7682423111&      -45.8823819291& 10.7328&   2.60793&  0.026263&    7529&  3.8499&   10188&  4.0525& 49.59&EB &$\delta$ Sct &       &LIA17,STR18\\	
           405605748&      202.0407789137&      -32.7236401436& 11.6890&   0.76493&  0.020972&    8129&  4.4030&    8421&  4.2182& 11.61&EB &$\delta$ Sct &     Y &\\	
           406421379&      308.2928216243&       24.8653789324& 11.3710&   2.02498&  0.014857&    8256&  4.2560&   10037&  4.2930& 23.44&EA &$\delta$ Sct &     Y &\\	
           406798603&      225.6920812427&       37.9101125343& 10.1329&   0.90639&  0.019095&    7987&  4.2135&    8432&  4.0935& 14.56&EB &$\delta$ Sct &       &KAH17,LIA17,ZXB15,DG15\\	
           406947878&      290.2984105960&       47.9785983500&  9.3918&   1.23147&  0.035547&    8068&  3.8697&    7699&  3.8499& 23.36&EB &$\delta$ Sct &  &SOU11,LEH13,KAH17,LIA17,MIS21\\	
           412638697&      329.6423921779&       60.9504944856& 11.2956&   0.98590&  0.022247&    8344&        &    7477&  3.9740& 10.80&EA &$\delta$ Sct &     Y &\\	
           413395842&      176.8449515428&      -23.0417568761&  8.8507&   0.89768&  0.027784&    7351&  4.1546&    7501&  4.0417&  9.42&EB &$\delta$ Sct &     Y &\\	
           428003183&      349.2207982107&       44.4884480054& 10.7131&   0.74260&  0.041583&    7345&  4.0477&        &        &      &EB &$\delta$ Sct &       &KAH22\\	
           431527184&      357.3813660223&       53.1346228524& 10.9091&   0.99850&  0.028854&    7591&  3.9536&    8154&  3.9327& 26.23&EB &$\delta$ Sct &       &KAH17,LIA17,KIM10\\	
           431555426&      357.6198613129&       50.6781736771&  9.6212&   1.77898&  0.427180&    7213&        &    7235&  3.8092& 13.26&EB &$\gamma$ Dor-$\delta$ Sct &     Y &\\	
           440003271&        2.5133397165&       46.3903078945&  7.6835&   2.63945&  0.064030&    8334&  4.1623&    9839&  4.2715& 27.47&EA &$\delta$ Sct &       &KAH22\\	
           440459327&       90.5398739179&       45.6903870005&  8.1075&   0.30233&  0.036479&    8394&  4.0445&    7935&  3.8429& 18.54&EW &$\delta$ Sct &     Y &\\	
           441006548&      210.0474875044&      -32.9364279291& 10.0813&   1.16720&  0.016255&    8386&  4.1924&    9792&  4.3133& 25.44&EB &$\delta$ Sct &     Y &\\	
           441406061&      310.8020240174&      -32.2931748296&  9.4763&   3.01802&  0.082078&    6916&        &    7019&  3.4860& 53.57&EB &$\delta$ Sct &       &SHI22\\	
           441598163&      115.2958520519&       76.0739139894& 10.2744&   3.30592&  0.066463&    7396&  3.5416&    7694&  3.6493& 40.72&EA &$\delta$ Sct &   &KAH17,LIA17,BRO74,KIM02,ROD10\\	
           447733367&      255.1928831684&      -64.0144302907& 11.8477&   2.63111&  0.108402&    6871&  3.8814&    6936&  3.9058& 11.19&EA &$\delta$ Sct &       &SHI22\\	
           449486448&      160.6955417028&      -72.9866185260& 10.4418&   3.51152&  0.630629&    7019&  3.7353&    7280&  3.7790& 18.06&EA &$\gamma$ Dor-$\delta$ Sct &     Y &\\	
           450089997&       72.1406335462&        8.5933140862&  9.2468&   2.46167&  0.132171&    6802&  3.4813&        &        &      &EB &$\delta$ Sct &     Y &\\	
           452734608&      195.4726275319&      -19.7746554648&  7.7217&   1.99393&  0.058240&    6111&        &    7032&  3.7233& 13.31&EB &$\delta$ Sct &       &BAR20\\	
           456609929&      282.3517441910&      -59.4205912111&  8.7194&   4.56750&  0.079275&    7643&  4.0974&    7763&  3.8448& 21.16&EA &$\delta$ Sct &     Y &\\	
           456905229&       26.2229199415&       19.8567993659&  8.0865&   1.69267&  0.065935&    7041&        &    6926&  3.9000& 11.26&EA &$\delta$ Sct &       &KAH22\\	
           459789367&      144.5279975768&       56.0186914005&  9.9765&   0.68737&  0.019501&    7664&  3.9623&    7593&  3.8172& 14.92&EB &$\delta$ Sct &       &KAH17,LIA17\\	
           468670361&       19.0875792722&       76.8027156233&  8.5340&   1.15206&  0.027899&    7824&  3.6188&    7201&  3.5652& 30.14&EB &$\delta$ Sct &     Y &\\	
           469979014&      342.2748751557&      -69.5671849817& 12.2478&   1.16259&  0.042609&    7679&  3.8584&    9431&  4.2337& 46.63&EB &$\delta$ Sct &     Y &\\	
           836353664&      132.1580534312&      -13.4350738321& 11.1811&   1.12319&  0.058210&    7526&        &    7355&  4.0598&  8.49&EB &$\delta$ Sct &     Y &\\	
           974128107&      183.7920872924&      -65.1783336508& 10.0375&   3.49309&  0.045772&        &        &        &        &      &EA &$\delta$ Sct &     Y &\\	 
\enddata
\tablecomments{Targe ID are sorted by increasing TIC number in the first column. The second and third columns present the equatorial coordinates in unit of degree. The fourth, seventh, and eighth columns are the TESS magnitudes, the effective temperature, and the surface gravity from the TIC input catalog (Stassun et al. 2019). The fifth and sixth columns are the orbital period and the dominant pulsation period. The ninth, tenth, and eleventh columns are the effective temperature, the surface gravity, and the luminosity from the Gaia DR3 (Gaia Collaboration 2022). "Ttype" column shows the type of binary system. "Ptype" column marks the type of pulsations. "New PB" denotes whether pulsations of the target are new detected. The references in the last column are provided below. }
\tablereferences{
 BAR20: Barcel{\'o} Forteza et al 2020, 
 BOW19: Bowman et al. 2019, 
 BRO74; Broglia \& Marin 1974, 
 CXH21: Chen et al. 2021, 
 CUI20: Cui et al. 2020, 
 DG15:  Do{\u{g}}ruel \& G{\"u}rol 2015,
 GAU13: Gaulme et al. 2013,
 GAU19: Gaulme \& Guzik 2019, 
 HAS21: Hasanzadeh et al. 2021, 
 HOL14: Holdsworth et al. 2014, 
 HON17: Hong et al. 2017,
 KAH17: Kahraman Ali{\c{c}}avu{\c{s}} et al. 2017, 
 KAH22: Kahraman Ali{\c{c}}avu{\c{s}} et al. 2022,
 KIM02: Kim et al. 2002, 
 KIM10: Kim et al. 2010,  
 LEE18: Lee et al. 2018, 
 LEE19: Lee et al. 2019, 
 LEH13: Lehmann et al. 2013, 
 LIA12: Liakos et al. 2012, 
 LIA17: Liakos \& Niarchos 2017, 
 LIA20: Liakos 2020, 
 MIS21: Miszuda et al. 2021, 
 MIS22: Miszuda et al. 2022, 
 MKR18: Mkrtichian et al. 2018, 
 RAP21: Rappaport et al. 2021, 
 ROD10: Rodr{\'\i}guez et al.2010,
 SHI22: Shi et al 2022,
 SOU11: Southworth et al. 2011, 
 SOU21: Southworth et al. 2021, 
 STR18: Streamer et al. 2018, 
 WK20:  Wang et al. 2020, 
 ZXB14, Zhang et al. 2014, 
 ZXB15: Zhang et al. 2015}
\end{deluxetable*}
\end{longrotatetable}

\begin{figure*}
\centering
\includegraphics[width=0.85\textwidth, angle = 0]{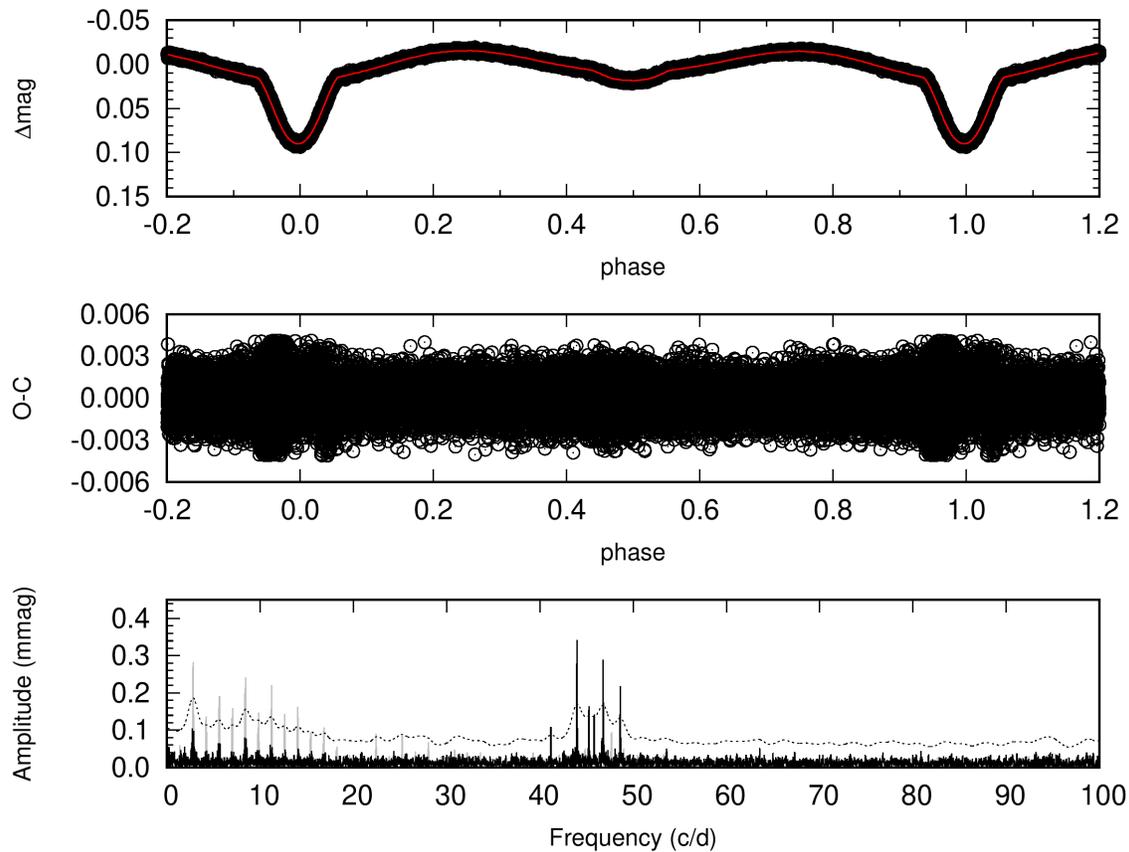}
  \caption{\label{Figure 1} Light curve,  O-C residuals, and amplitude spectrum of TIC 125498216.  The light curve is displayed with black circles in the top panel, the O-C residuals is presented in the middle panel,  and the amplitude spectra is depicted in the bottom panel. The red line in the top panel denotes the synthesis mean curve. The peaks marked in gray are possible orbital-harmonics. The dashed line marks the positions of S/N =5.}
\end{figure*}

\begin{figure*}
\centering
\includegraphics[width=0.8\textwidth, angle = 0]{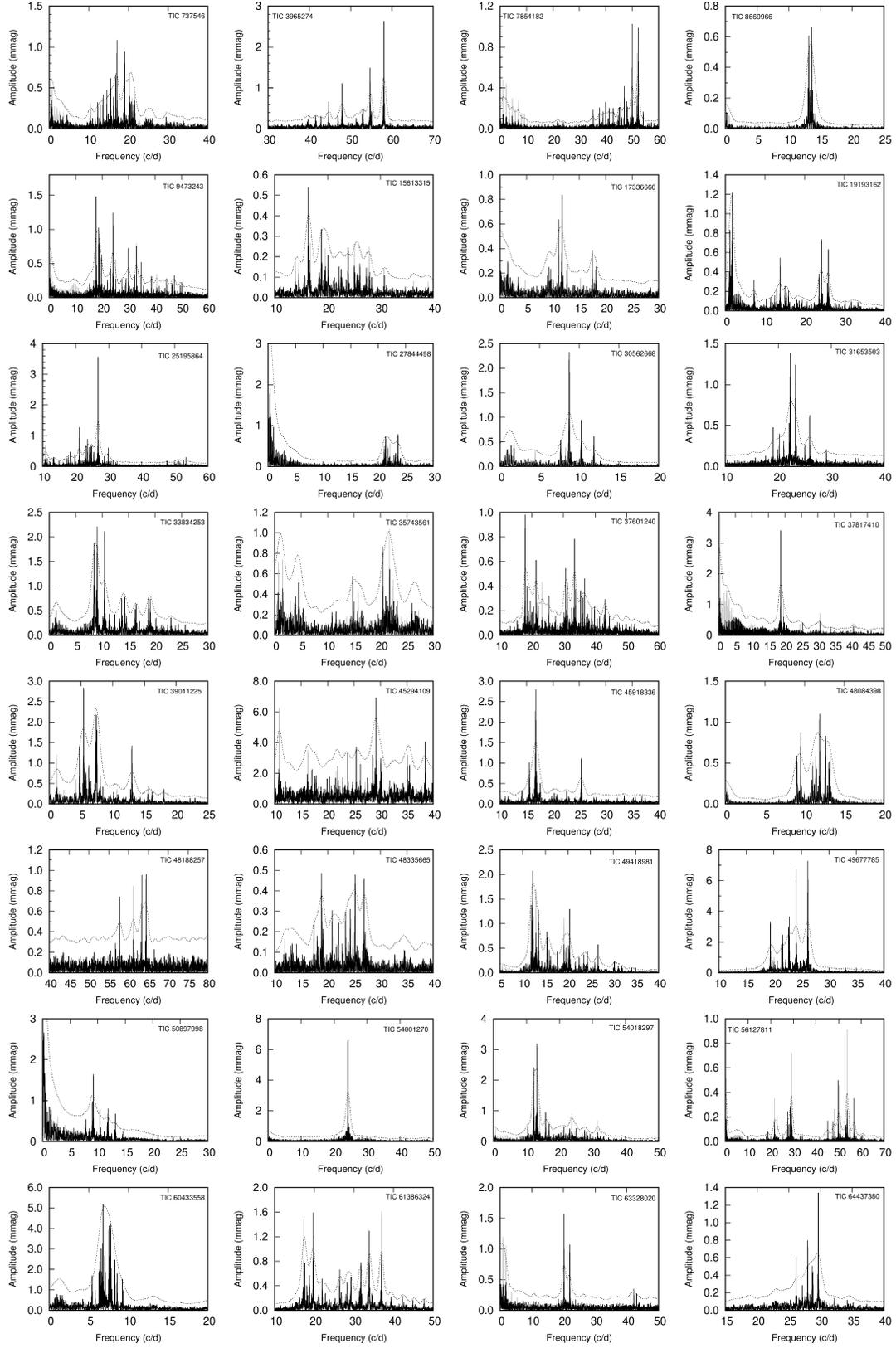}
  \caption{\label{Figure 2} Amplitude spectra of the pulsating components in eclipsing binaries. The dashed lines mark the positions of S/N =5. All the component figures are available in the Appendix.}
\end{figure*}
\begin{figure*}
\centering
\includegraphics[width=0.85\textwidth, angle = 0]{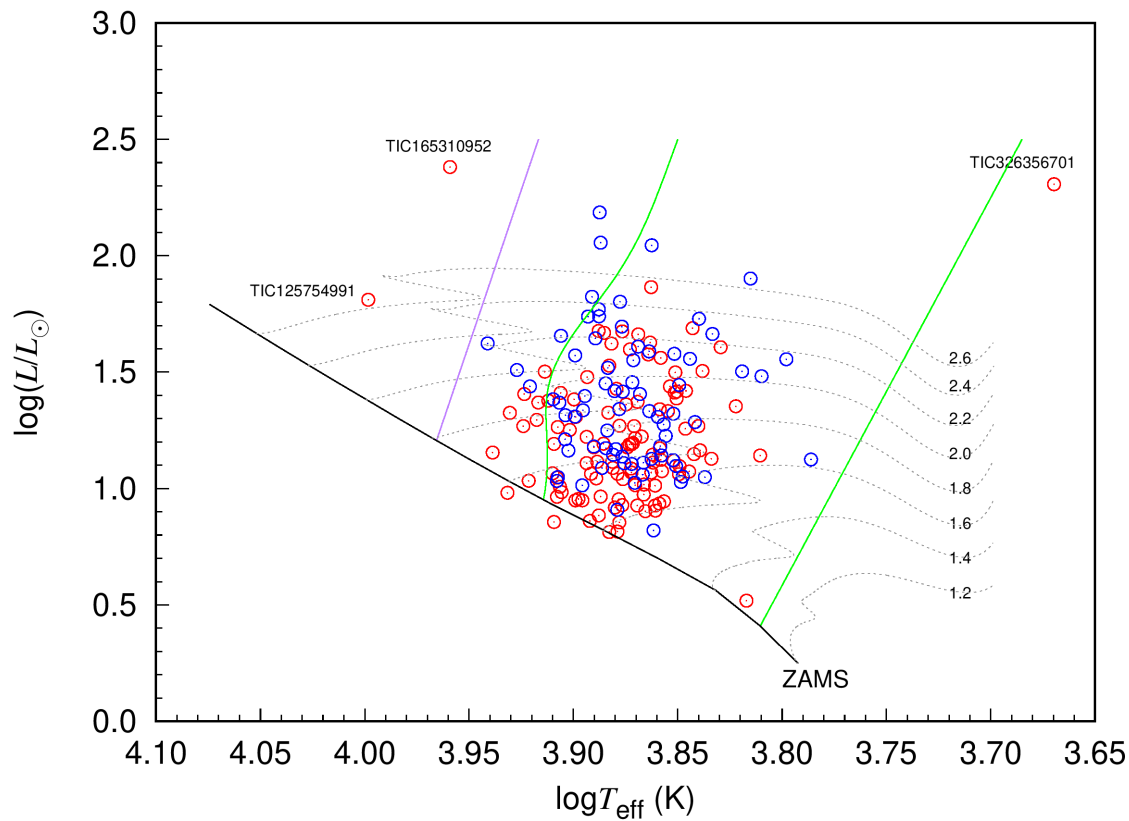}
  \caption{\label{Figure 3} The distribution of the targets in the Hertzsprung-Russell diagram. Newly discovered and old targets are marked with red and blue circles, respectively. The ZAMS line and the evolutionary tracks of eight different initial stellar mass with $Z$ = 0.015 are computed with the one-dimensional stellar evolution code Modules for Experiments in Stellar Astrophysics (MESA,  Paxton et al. 2011,  2013,  2015,  2018) using the same input physics with Chen et al. (2021).  The green lines represent the instability strip derived by Xiong et al. (2016) and the purple line is the theoretical blue edge of the $\delta$ Scuti instability strip derived by Li \& Stix (1994). Luminosities $L$ are taken from Gaia DR3, while the values of $T_{\rm eff}$ of the stars are taken from the TESS input catalog.}
\end{figure*}
\begin{figure*}
\centering
\includegraphics[width=0.85\textwidth, angle = 0]{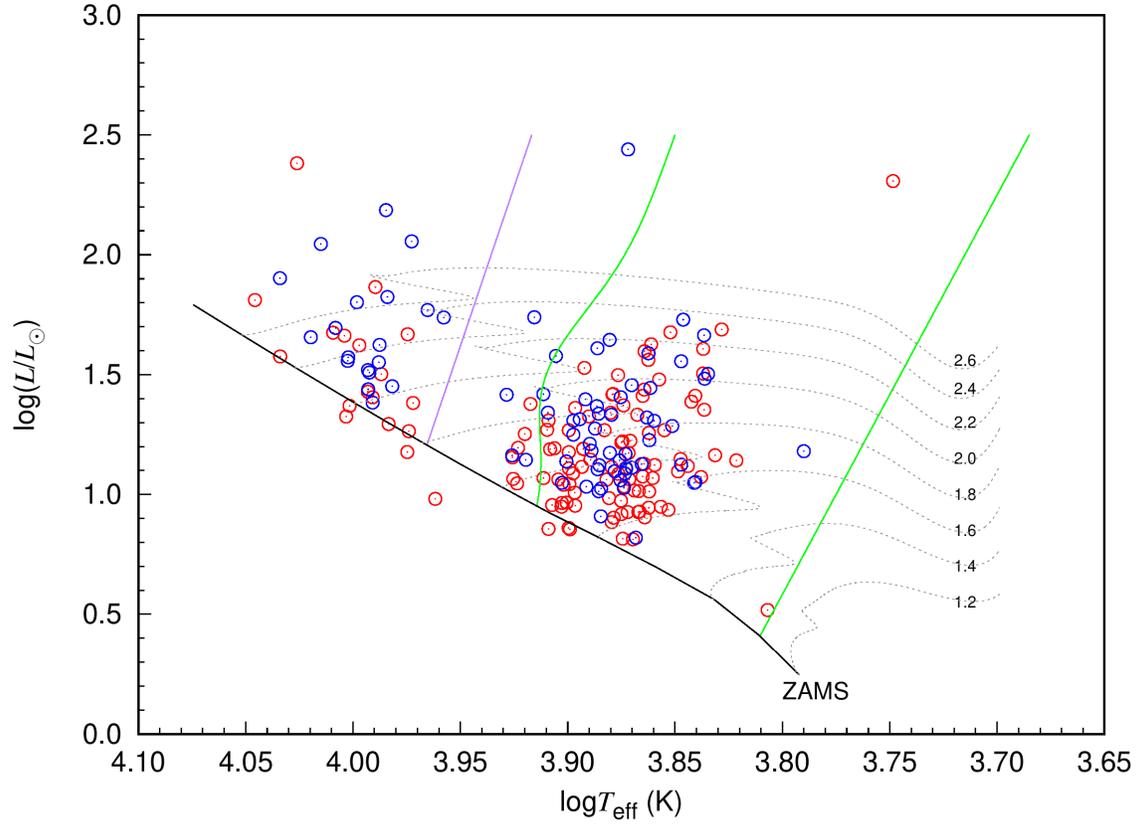}
  \caption{\label{Figure 4} The same as in Figure 3,  but the values of effective temperature $T_{\rm eff}$ of the stars are taken from the Gaia DR3.}
\end{figure*}

\begin{figure*}
\centering
\includegraphics[width=0.85\textwidth, angle = 0]{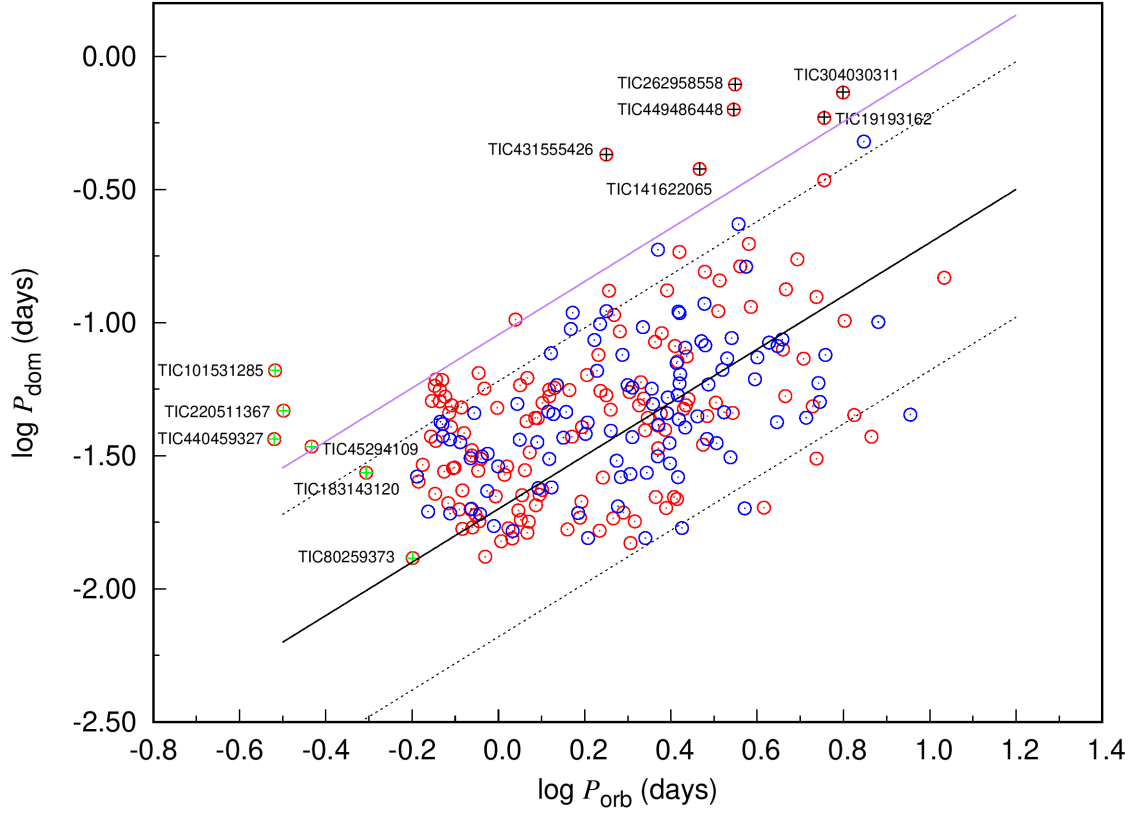}
  \caption{\label{Figure 5} Distribution of the pulsating stars in eclipsing binaries in the $P_{\rm dom}$-$P_{\rm orb}$ diagram.  Newly discovered and old targets are marked with red and blue circles, respectively. The solid line in black is the theoretical relation  by Zhang et al.  (2013),  and the two dashed lines correspond to its 3-$\sigma$ uncertainties.  The purple line denotes the upper limit of the $P_{\rm dom}$/$P_{\rm orb}$ ratio derived by Zhang et al.  (2013) for the $\delta$ Scuti stars in eclipsing binaries. The systems marked with green and black cross symbols correspond to targets in Figure 6 and 7, respectively.} 
\end{figure*}
\begin{figure*}
\centering
\includegraphics[width=0.85\textwidth, angle = 0]{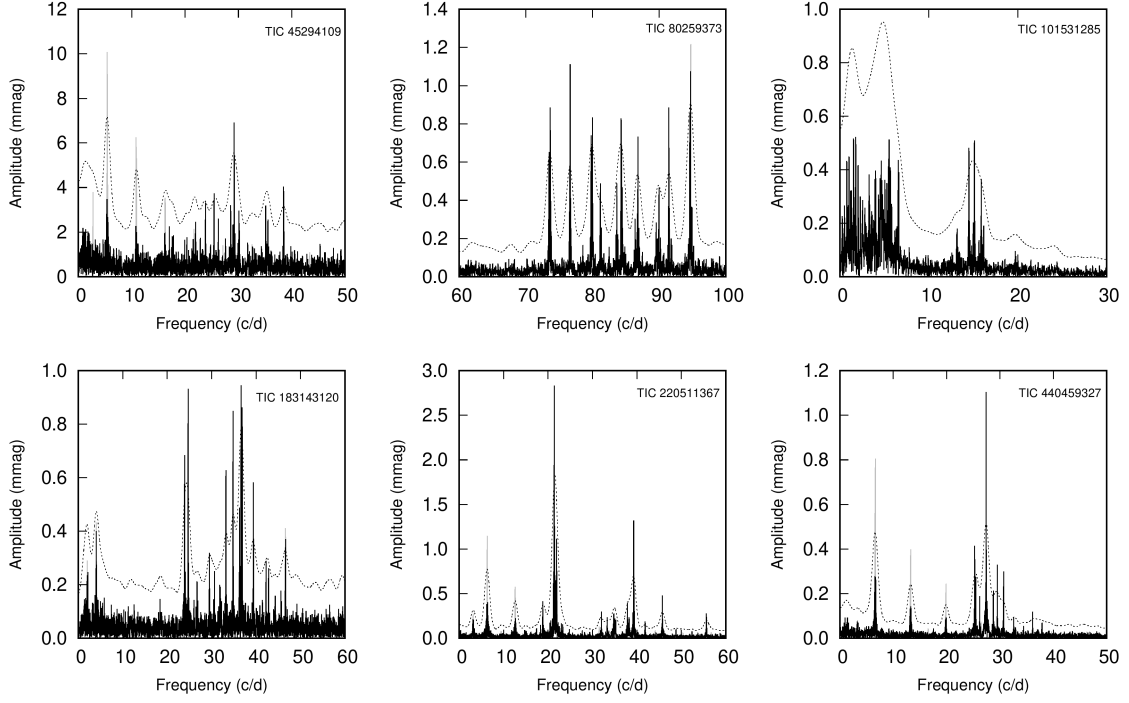}
  \caption{\label{Figure 6} Fourier amplitude spectra for the pulsating components in eclipsing binaries with $P_{\rm orb}$ $<$ 0.65 days. The dashed lines mark the positions of S/N =5.}
\end{figure*}
\begin{figure*}
\centering
\includegraphics[width=0.85\textwidth, angle = 0]{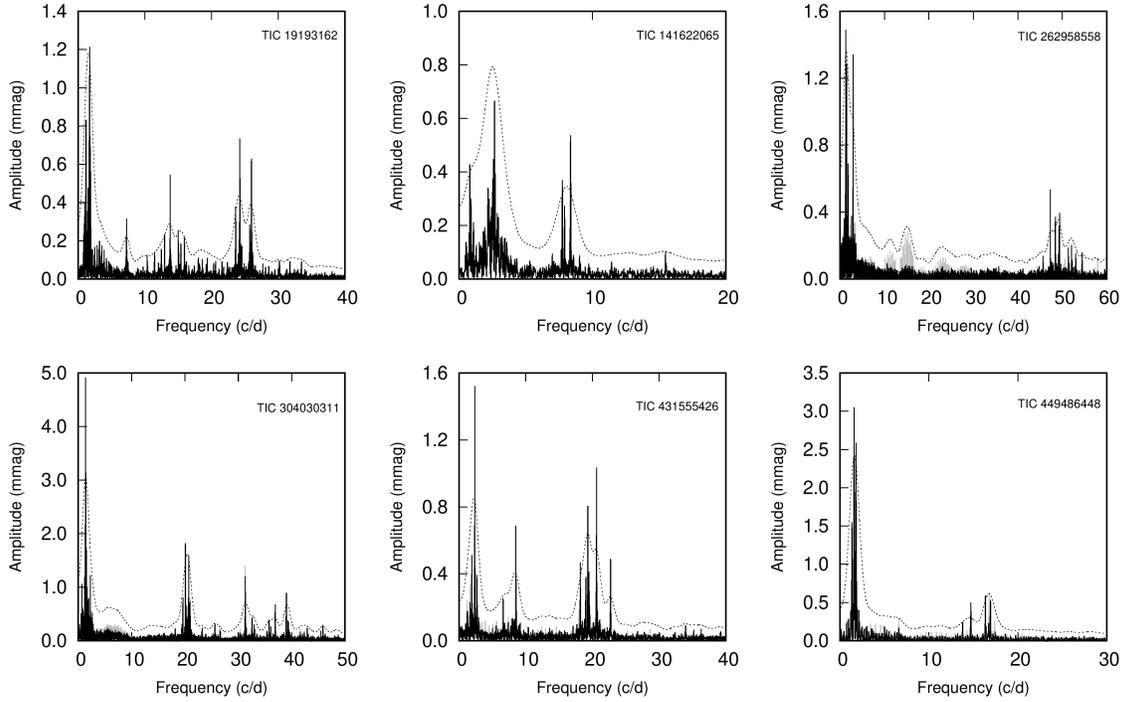}
  \caption{\label{Figure 7} Fourier amplitude spectra of objects TIC 19193162,  TIC 141622065,  TIC 262958558,  TIC 304030311,  TIC 431555426,  and TIC 449486448, respectively. The dashed lines mark the positions of S/N =5.}
\end{figure*}
\begin{figure*}
\centering
\includegraphics[width=0.85\textwidth, angle = 0]{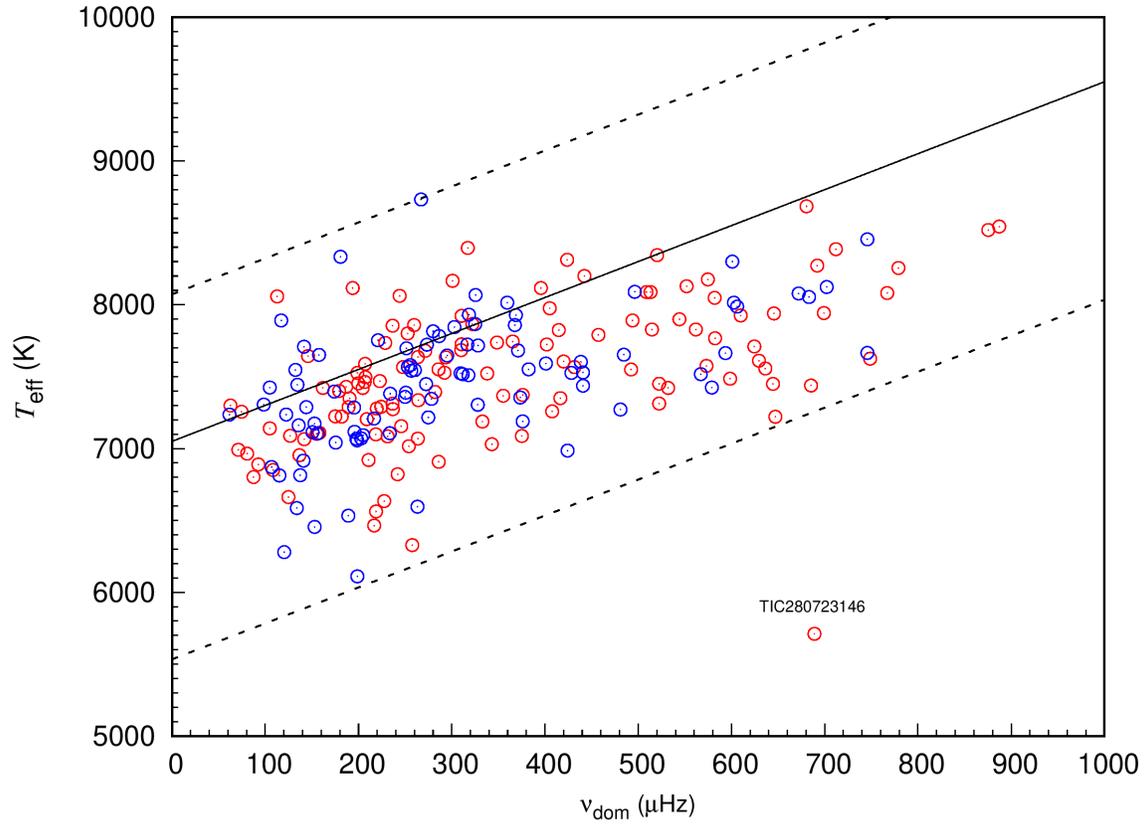}
  \caption{\label{Figure 8} Distribution of the $\delta$ Scuti-type pulsating stars in eclipsing binaries within the $T_{\rm eff}$-$\nu_{\rm dom}$ relation. Newly discovered and old targets are marked with red and blue circles, respectively.  The black solid line marks the theoretical relation derived by Barcel{\'o} Forteza et al.(2020),  and the dashed lines denote the limits of their predicted dispersion due to the gravity-darkening effect.}
\end{figure*}
\begin{figure*}
\centering
\includegraphics[width=0.85\textwidth, angle = 0]{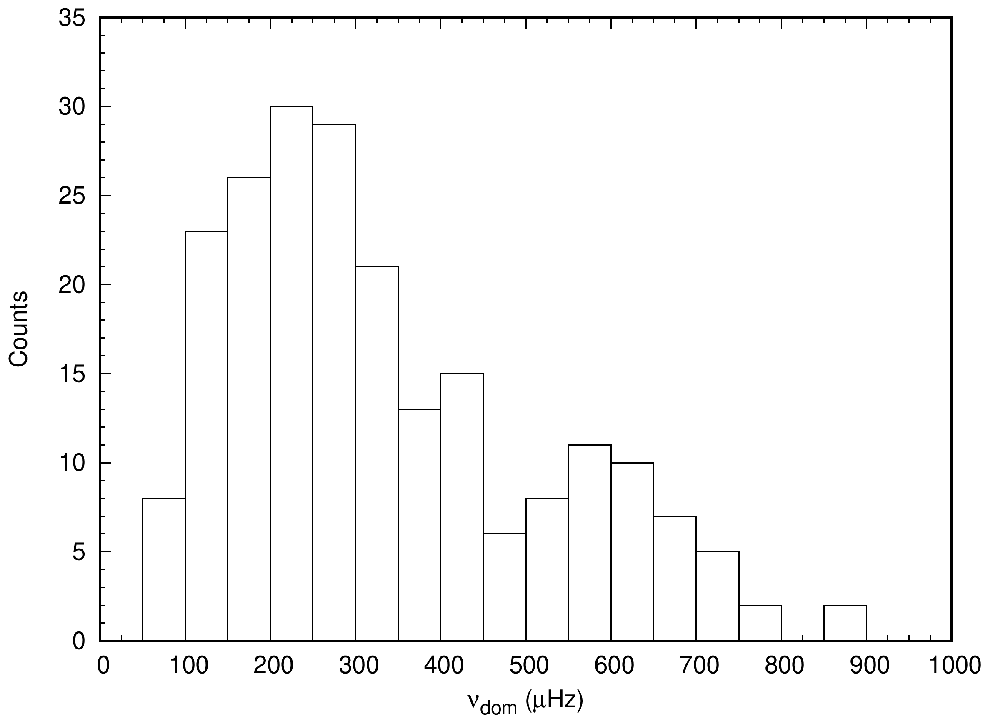}
  \caption{\label{Figure 9} Distribution of the dominant pulsation frequencies of $\delta$ Scuti pulsators in eclipsing binaries.}
\end{figure*}

\appendix
\section{Supplementary figures of amplitude spectra for the 242 targets.}
\setcounter{figure}{0}
\renewcommand{\figurename}{Supplementary Figure}
\begin{figure*}
\centering
\setcounter {figure}{0}
\includegraphics[width=0.8\textwidth, angle = 0]{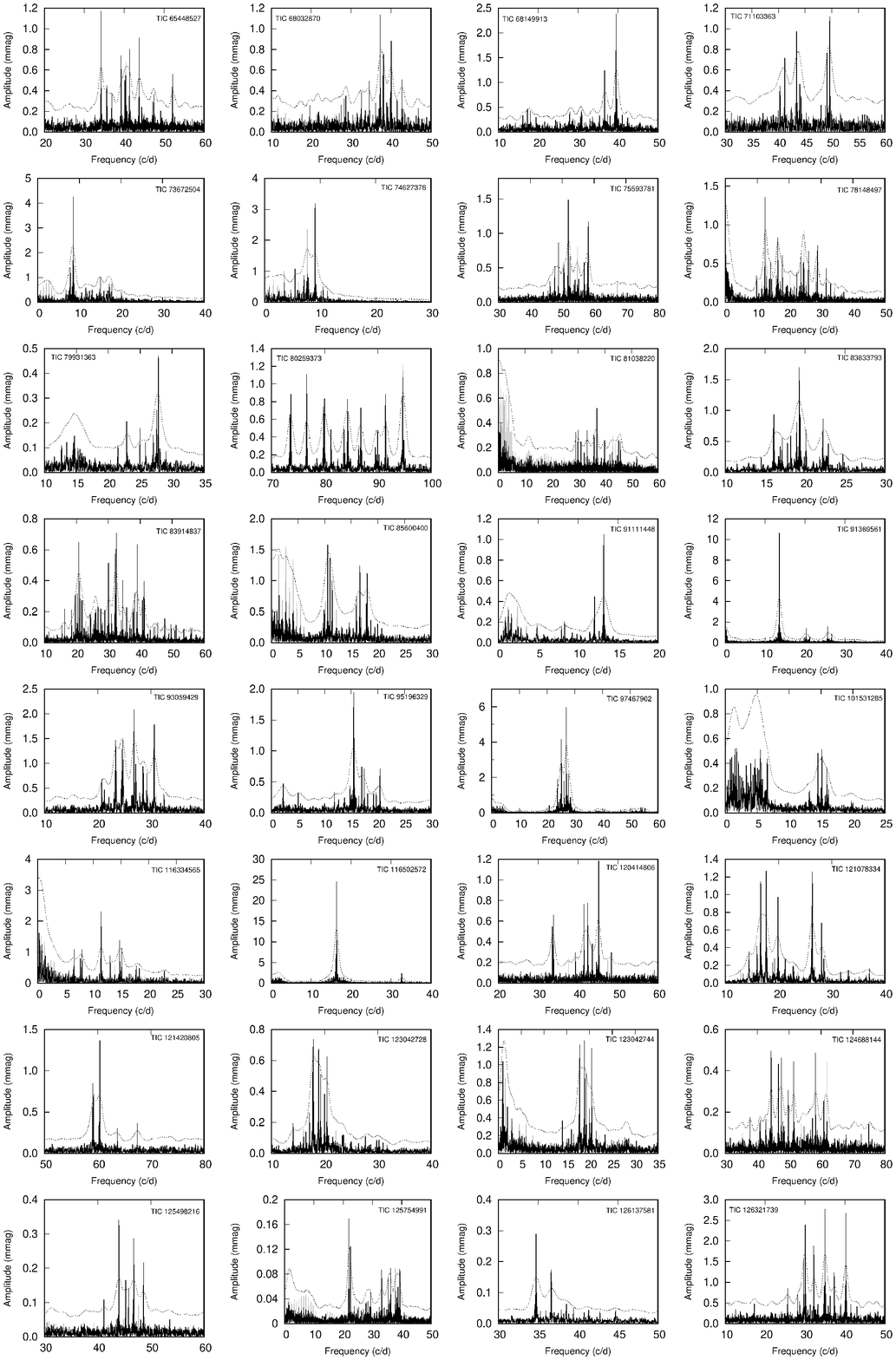}
  \caption{\label{Figure S1}  Continued to Figure 2 for different targets.}
\end{figure*}
\begin{figure*}
\centering
\setcounter {figure} {0}
\includegraphics[width=0.8\textwidth, angle = 0]{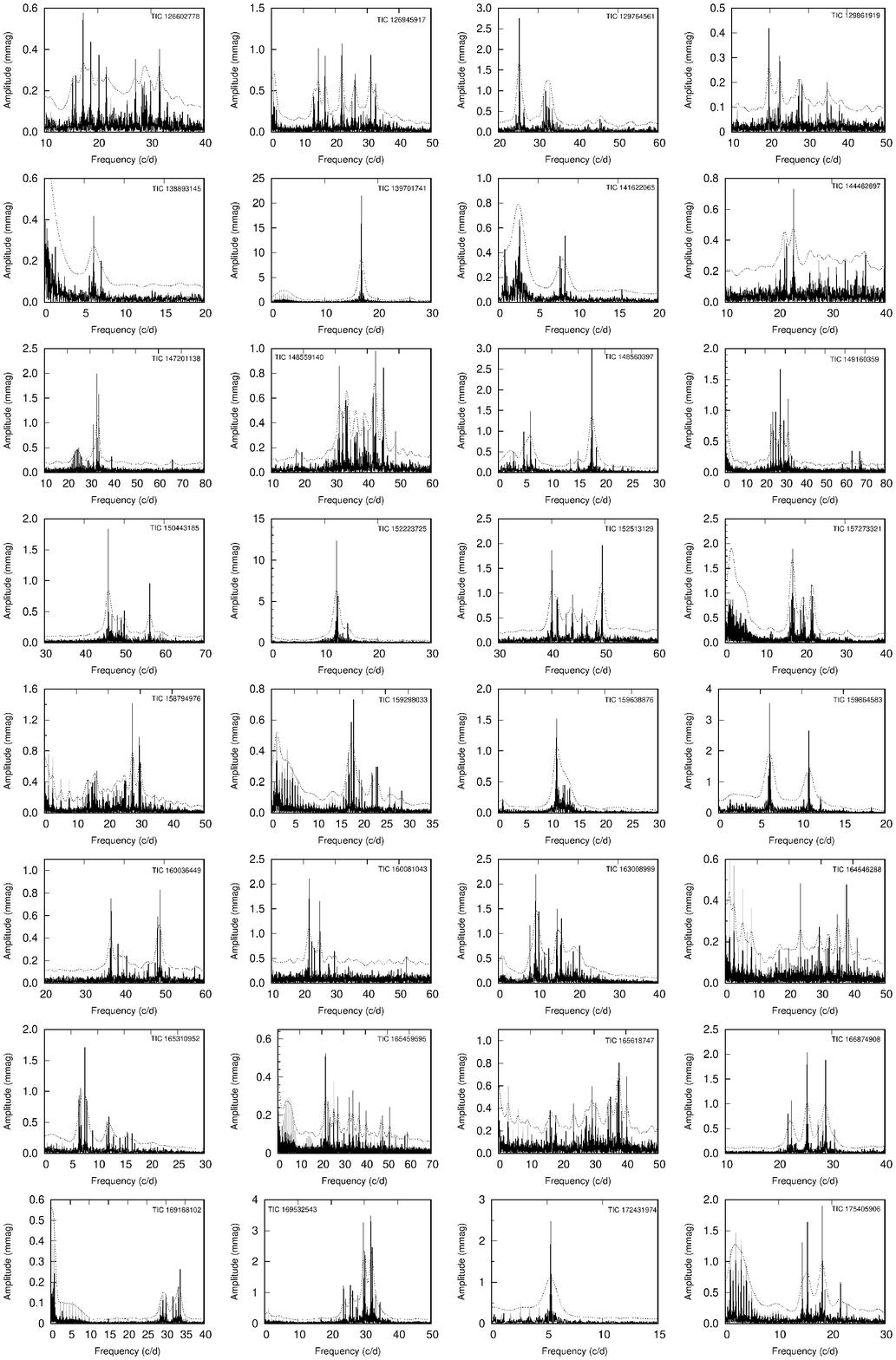}
  \caption{\label{Figure S2}  (continued) }
\end{figure*}
\begin{figure*}
\setcounter {figure} {0}
\centering
\includegraphics[width=0.8\textwidth, angle = 0]{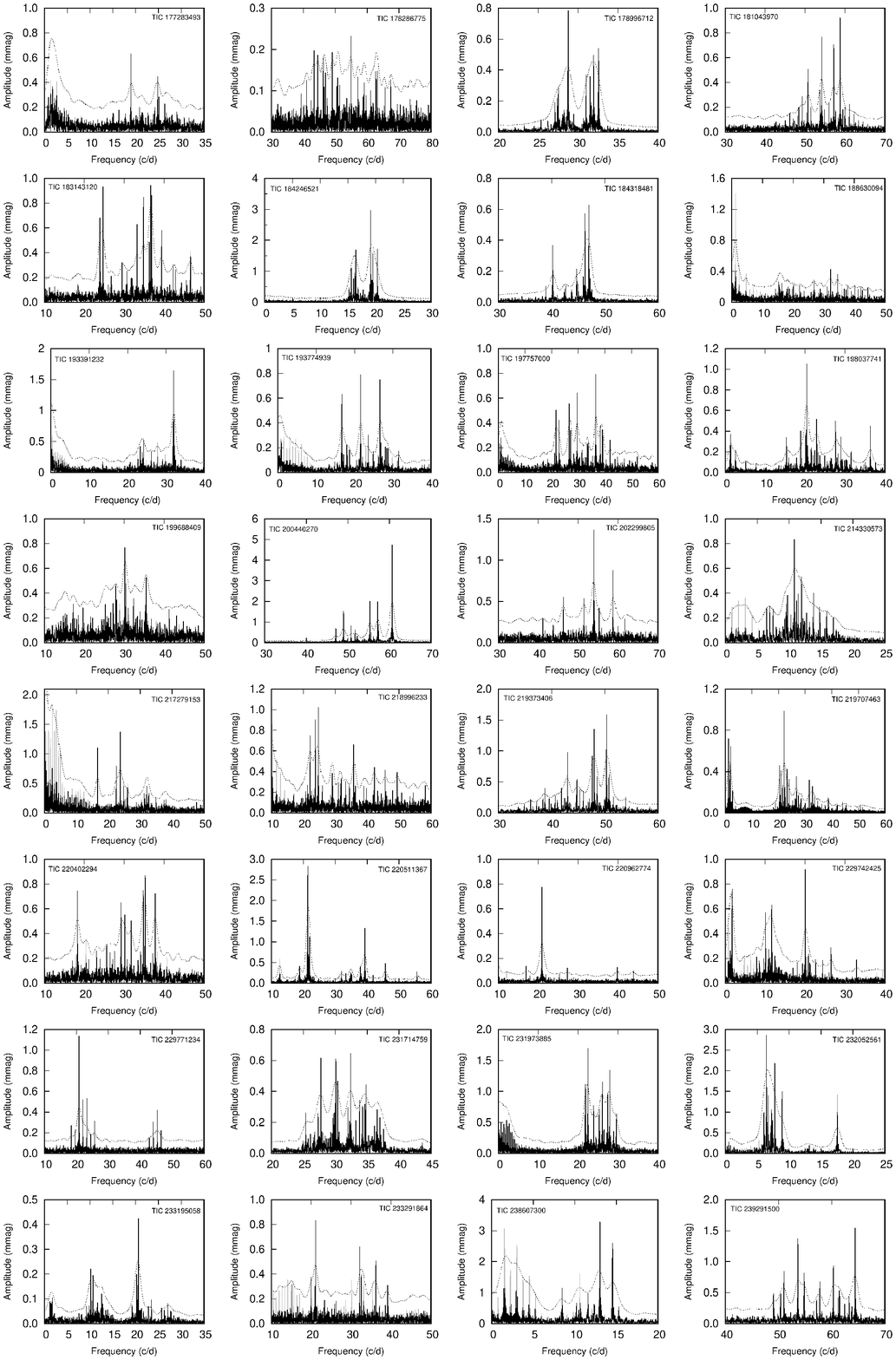}
  \caption{\label{Figure S3}  (continued) }
\end{figure*}
\begin{figure*}
\setcounter {figure} {0}
\centering
\includegraphics[width=0.8\textwidth, angle = 0]{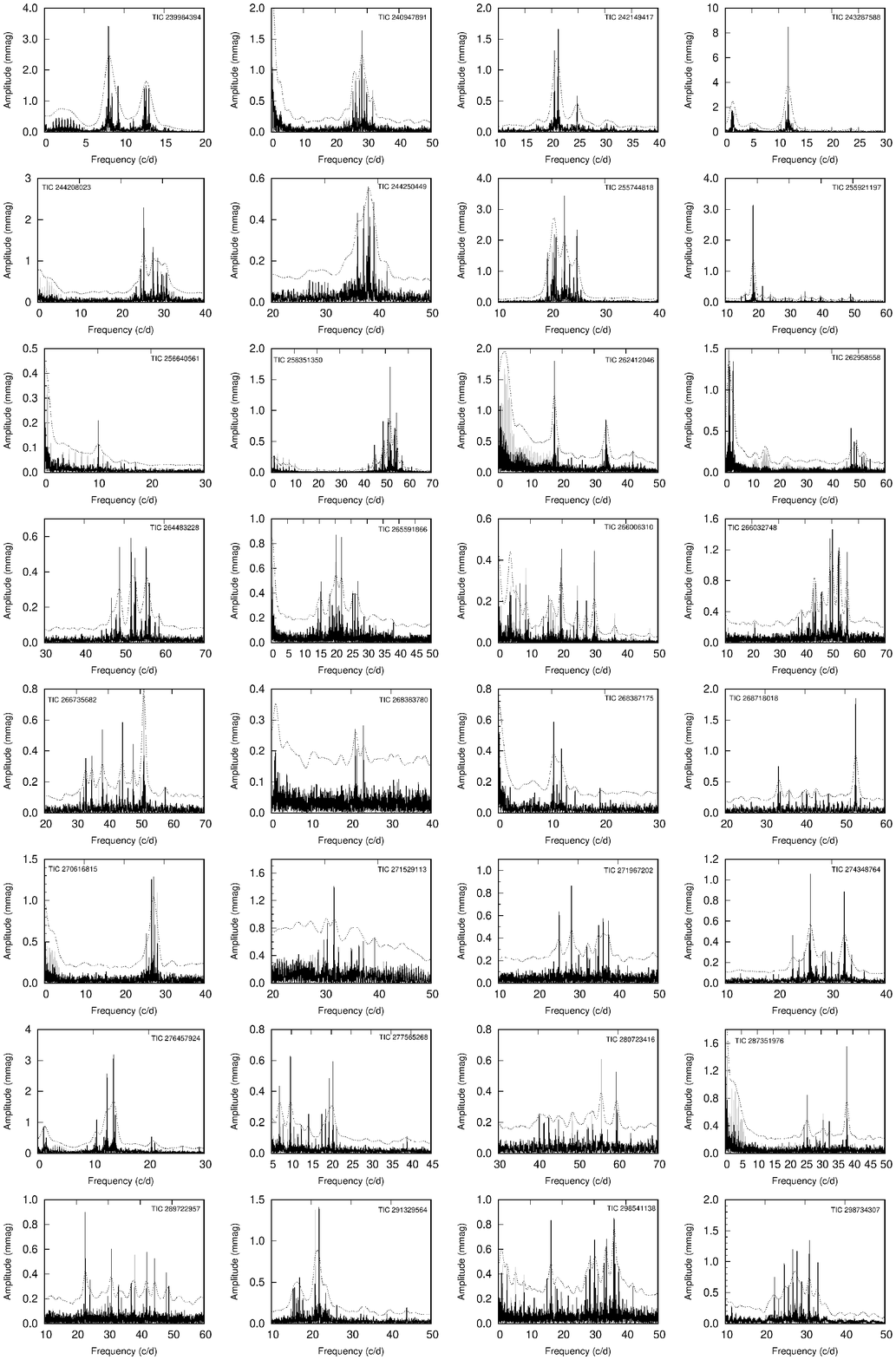}
  \caption{\label{Figure S4}  (continued)  }
\end{figure*}
\begin{figure*}
\setcounter {figure} {0}
\centering
\includegraphics[width=0.8\textwidth, angle = 0]{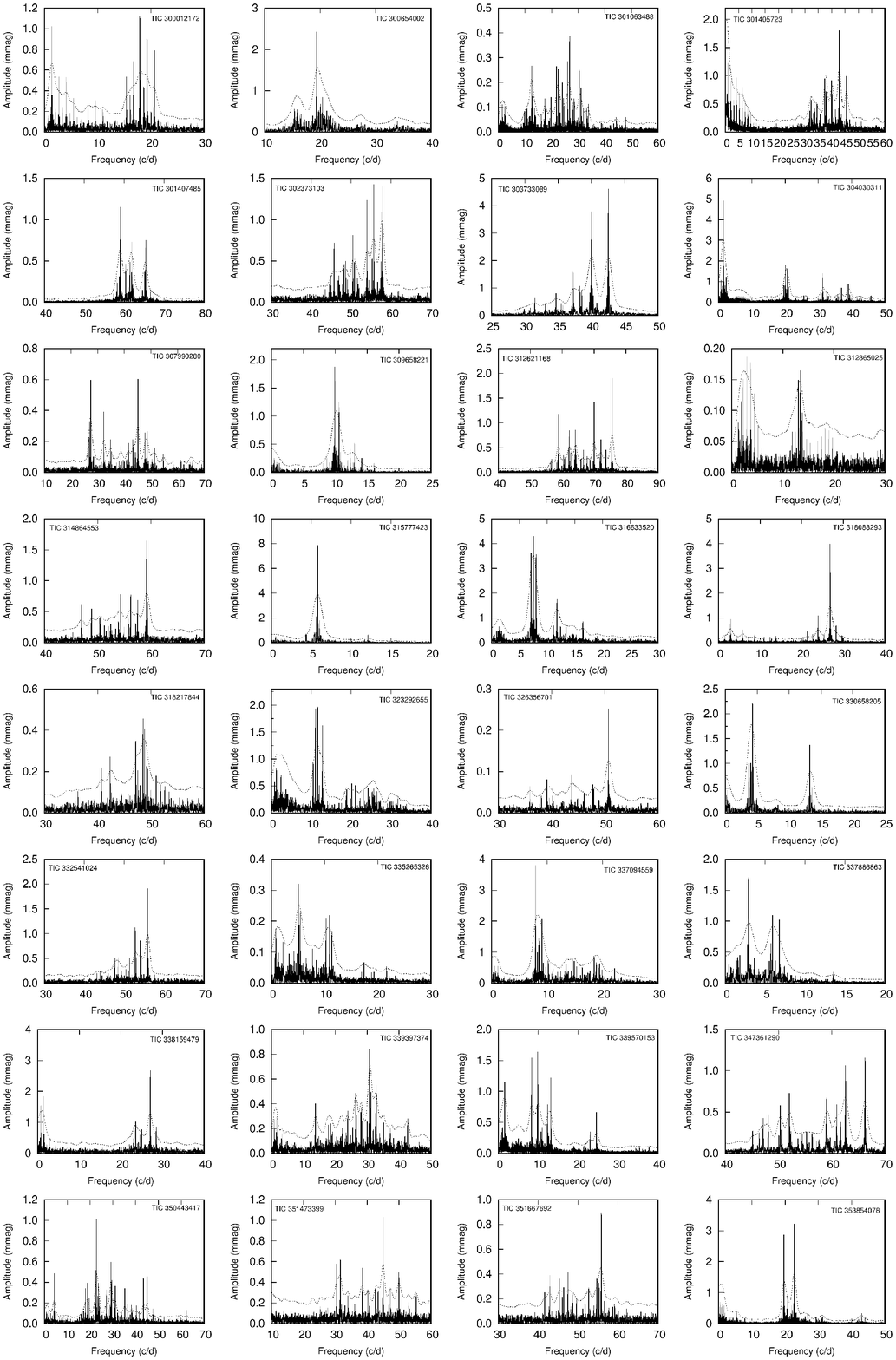}
  \caption{\label{Figure S5}  (continued) }
\end{figure*}
\begin{figure*}
\setcounter {figure} {0}
\centering
\includegraphics[width=0.8\textwidth, angle = 0]{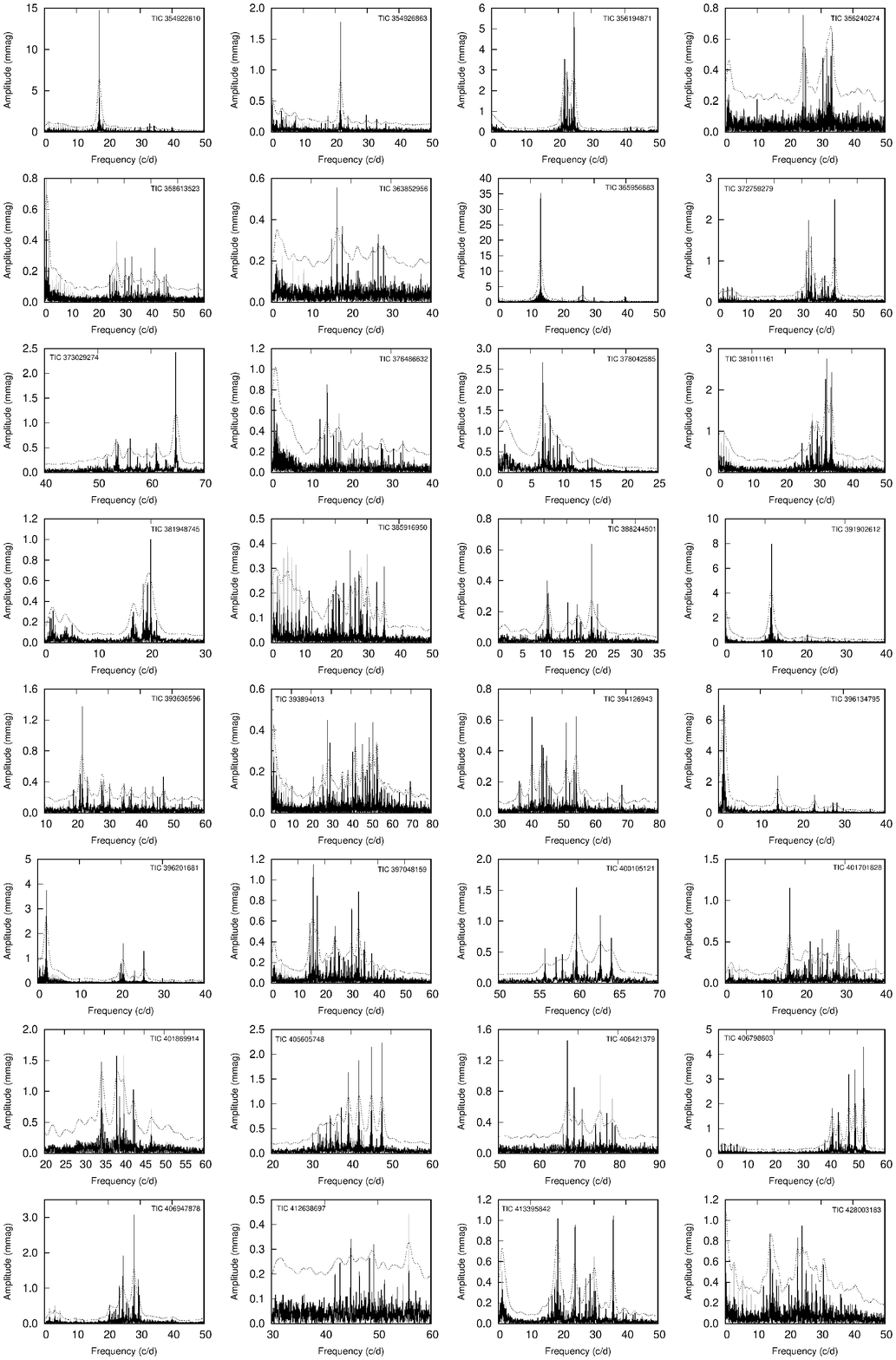}
  \caption{\label{Figure S6} (continued)}
\end{figure*}
\begin{figure*}
\setcounter {figure} {0}
\centering
\includegraphics[width=0.8\textwidth, angle = 0]{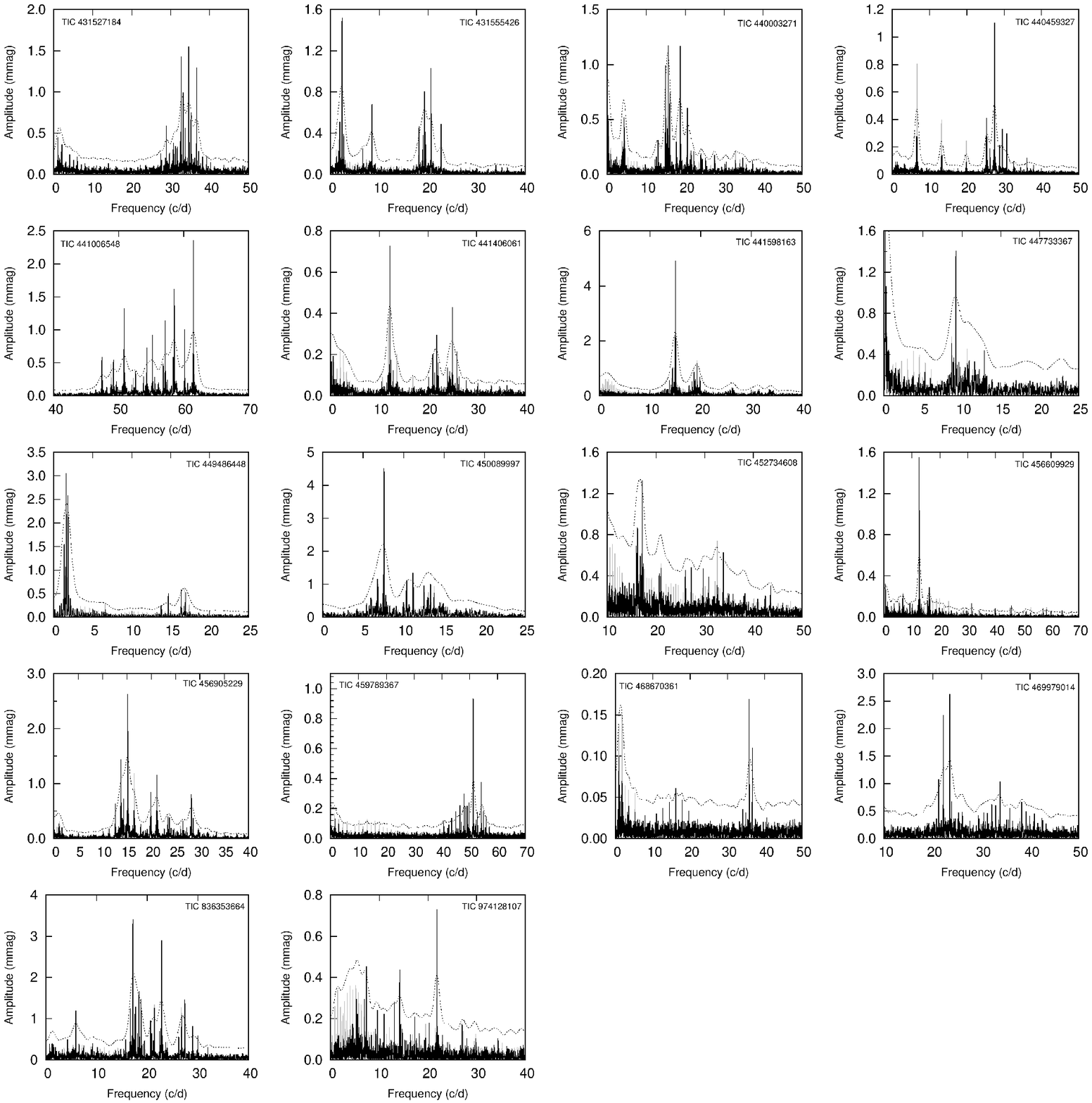}
  \caption{\label{Figure S7} (continued)}
\end{figure*}


\end{document}